\documentclass[onecolumn,superscriptaddress,secnumarabic,amssymb,amsmath,nobibnotes,aps,prd,nofootinbib,altaffilletter]{revtex4}
\usepackage{amsmath}
\usepackage{amsfonts,color}
\usepackage{amssymb,float}
\usepackage[utf8]{inputenc}
\usepackage{color}
\usepackage{verbatim}
\usepackage[unicode=true,pdfusetitle,
 bookmarks=true,bookmarksnumbered=true,bookmarksopen=true,bookmarksopenlevel=3,
 breaklinks=false,pdfborder={0 0 1},backref=false,colorlinks=true]
 {hyperref}
\usepackage{enumitem}
\usepackage{graphicx}
\usepackage{subfig}
\usepackage{verbatim} 
\usepackage{appendix}
\usepackage{hyperref}
\usepackage[most]{tcolorbox}
\usepackage{orcidlink}
\usepackage{ulem}

\makeatletter
\long\def\dddddot#1{%
  {\mathop {#1}\limits ^{\vbox to-1.4\ex@ {\kern -\tw@ \ex@ \hbox {\normalfont .....}\vss }}}%
}
\long\def\multidots#1#2{%
  \count@=0
  {{\mathop {#2}\limits ^{\vbox to-1.4\ex@ {\kern -\tw@ \ex@ \hbox {\normalfont %
  \loop%
  \ifnum#1>\count@%
  .%
  \advance\count@ by1%
  \repeat%
  }\vss }}}}%
}
\makeatother

\newcommand{\dut}[3]{#1_{#2}^{\phantom{#2}#3}}

\begin{document}

\title{Non-fluid like Boltzmann code architecture for early times \texorpdfstring{$f(T)$}{} cosmologies}

\author{Aram Aguilar}
\email{armandoaram$\_$94@hotmail.com}
\affiliation{Instituto de Ciencias Nucleares, Universidad Nacional Aut\'{o}noma de M\'{e}xico, 
Circuito Exterior C.U., A.P. 70-543, M\'exico D.F. 04510, M\'{e}xico.}

\author{Celia Escamilla-Rivera\orcidlink{0000-0002-8929-250X}}
\email{celia.escamilla@nucleares.unam.mx}
\affiliation{Instituto de Ciencias Nucleares, Universidad Nacional Aut\'{o}noma de M\'{e}xico, 
Circuito Exterior C.U., A.P. 70-543, M\'exico D.F. 04510, M\'{e}xico.}

\author{Jackson Levi Said\orcidlink{0000-0002-7835-4365}}
\email{jackson.said@um.edu.mt}
\affiliation{Institute of Space Sciences and Astronomy, University of Malta, Malta, MSD 2080}
\affiliation{Department of Physics, University of Malta, Malta}

\author{Jurgen Mifsud}
\email{jurgen.mifsud@um.edu.m}
\affiliation{Institute of Space Sciences and Astronomy, University of Malta, Msida, MSD 2080, Malta}
\affiliation{Department of Physics, University of Malta, Msida, MSD 2080, Malta}

\begin{abstract}
There have been several works that have studied scalar cosmological perturbations in $f(T)$ teleparallel gravity
theories to understand early cosmic times dynamics. In this direction, the perturbations presented
have been performed by considering $f(T)$ extensions as an effective fluid-like scheme, where the equation-of-state contains extra terms due to the torsion. 
In this work, we discuss introducing a non-fluid-like approach as a direct consequence of $f(T)$ extensions, particularly for $f(T)$ power law model scenarios. This approach will be compared using CMB constraints data from Planck 2018 and SDSS catalogs, showing a change in about 17\% in $C_{l}$ at $l< 10^{1}$ from the ones reported in the literature as a fluid-like approach, which will bring significant changes in the analysis on cosmological tensions at early cosmic times. 
\end{abstract}

\maketitle

\section{Introduction}
\label{sec:Intro}

In the last few years, there have been several attempts to explain the accelerated expansion of the Universe. The most known of these is the one offered by the standard cosmological model, the so-called $\Lambda$CDM model, which assumes General Relativity (GR) as the theory which describes gravity, that spacetime at large scales is homogeneous and isotropic, and that dark matter and dark energy components permeate on large scales. Nevertheless, the absence of direct evidence of dark matter and dark energy has led to a reevaluation of this model. Also, the growing disparity
between the predicted cosmological parameter $H_0$, this is, the Hubble parameter evaluated at present time, deduced from the early universe and the one observed from local measurements, which is one of the most important cosmological issues called the $H_0$ tension, has been a source of reconsideration of the $\Lambda$CDM model \cite{Abdalla:2022yfr}, together with other anomalies such as the value of $\sigma_{8}$ and $\Omega_{k}\neq 0$.
To solve these issues, new proposals have emerged in the context of alternative theories of gravity \cite{Clifton:2011jh}, where this dark sector can emerge as a result of extra terms in modified gravitational field equations. Furthermore, cosmological perturbations in GR and alternative versions of theories of gravity have been thoroughly studied. At linear order, cosmological perturbations in GR and alternative theories of gravity can be useful to describe the dark sector effects, including structure formation, through inhomogeneity and anisotropy. However, the effects of some models can not be seen at this order, for it is necessary to expand up to major order to see these contributions \cite{Nakamura:2020pre}. 

Some of the proposals that have gained strength over the last years are extensions of the so-called teleparallel equivalent of general relativity (TEGR) \cite{Bahamonde:2021gfp} which, instead of curvature, relies on torsion to explain gravity \cite{Formiga:2020wsk}. Extensions of TEGR have resulted to be relevant and interesting when studying some cosmological scenarios. Inspired by the $f(R)$ extension of Einstein-Hilbert action, the simplest generalization of TEGR is $f(T)$ teleparallel gravity, where the Lagrangian density becomes an arbitrary function of $T$. In comparison to $f(R)$ gravity, whose fourth-order field equations might lead to pathologies, $f(T)$ gravity has the remarkable feature of being of second-order in its field equations \cite{Cai:2015emx}, where there are two dynamical variables; the tetrad field, which carries the gravitational effects information, and the spin connection, which is a Lorentz degree of freedom and, therefore, carries the inertial effects information. Also, $f(T)$ gravity seems to be consistent with solar system tests \cite{Farrugia:2016xcw} and with other astronomical data \cite{Nunes:2016qyp, Qi:2017xzl}, making it a viable alternative candidate to the $\Lambda$CDM cosmology. In this landscape, a $f(T)$ power law model has been proven to be a good candidate since it produces the same effect as dark energy and exhibits phases dominated by matter and radiation \cite{Bengochea:2008gz}. Also, it does not exclude the $\Lambda$CDM paradigm and is in excellent agreement with current experimental bounds \cite{LeviSaid:2020mbb}. In this paper, we will focus on this model.

Furthermore, cosmic acceleration can be addressed in the early universe through cosmological perturbations \cite{Bucher:2015eia}, in particular from scalar perturbations since these are related to structure formation and the dark sector, whereas the tensor perturbations are related to gravitational waves; in \cite{Heisenberg:2023tho} a recent study in general teleparallel gravity (using torsion and non-metricity) gravity was performed by doing tensor perturbations, and this study was extended to $f(T)$, $f(Q)$ and $f(G)$ gravity. Regarding the vector perturbations, it is well known that they decay extremely fast so their effects are null at the present time. Several works in $f(T)$ scalar perturbations have been treated \cite{Nunes:2018xbm, Zheng:2010am, Benetti:2020hxp, Chen:2010va, Li:2011wu, Heisenberg:2023wgk, Duchaniya:2023aeu, Capozziello:2023giq}. One particular similarity between them is that an effective fluid-like approach is considered. The fluid-like approach consists of assuming a standard primordial perturbation pattern for a perfect fluid; so, instead of directly perturbing the field equations it is associated, at background level, geometric terms to the energy-momentum tensor. These geometric terms come from the modified theory of gravity and allow the definition of an effective equation-of-state which enables to mimicking of the linear perturbations behaviour with a certain precision \cite{Benetti:2020hxp}. However, this approach might not be able to obtain the complete perturbation information, because the dynamical field equations remain unaltered; the extra geometric terms are added to the matter sector, so the standard form of the field equations is preserved. In this work, we are going to present a second approach that consists of a non-fluid-like scheme, which can be considered as the canonical procedure since it directly perturbs the basic tensor elements and evolves the quantities to obtain the perturbed field equations (Boltzmann equations), from which you can obtain the desired cosmic dynamics that is well known. This approach can be seen as complete in the sense that we are fully recovering the perturbation information due to the perturbed field equations, furthermore, at background, a fluid-like approach is well-behaved but at perturbative level we expect these two approaches to differ due to the structure of fields equations.

In this paper, we have two main goals. Firstly, we will develop the $f(T)$ scalar perturbed field equations, obtained in the non-fluid-like scheme, for the Boltzmann solver 
Secondly, we will use these perturbed equations to study the power spectra for $f(T)$ cosmic models within a non-fluid-like approach. In particular, we will focus on describing the code architecture to implement the $f(T)$ power law model derived from TEGR theories.
The code methodology introduces new changes on certain modules that were adapted with the scalar perturbations for the $f(T)$ cosmologies. Also, as an extension, we compute the corresponding power spectra to show the differences in comparison to fluid-like approaches most used in the literature. This analysis will include recent CMB observational catalogs.

The outline of this work is organized as follows: In Sec.~\ref{sec:TG} we will explain the foundations of teleparallel gravity (TG);
in Sec.~\ref{subsec:Background}, we present the associated mathematical tools to develop the cosmological models in TG, while in Sec.~\ref{subsec:f(T)-background}, we describe the $f(T)$ model as a direct extension of the TEGR case. In  Sec.~\ref{sec:perturbfT} we discuss how the non-fluid-like scheme can be connected to cosmological scalar perturbations. Also, we will discuss the evolution equations associated with this theory. In Sec.~\ref{sec:BC} we treat carefully the description and implementation of the Boltzmann code architecture adapted to our non-fluid-like approach and we discuss the power spectra for $f(T)$ power law cosmologies using recent CMB data from Planck and Sloan Digital Sky Survey (SDSS) catalogs, including data from the integrated Sachs-Wolfe effect. Finally, in Sec.~\ref{sec:conclusions} we review our main conclusions.

\section{Teleparallel gravity and its \texorpdfstring{$f(T)$}{fT} extension}
\label{sec:TG}

In this section, we first introduce the general layout of TG and its geometric formalism. We then develop the general class $f(T)$ cosmology at background level.


\subsection{Background and main quantities}
\label{subsec:Background}

TG is a particular case theory within the metric affine geometry \cite{Bahamonde_2023} where torsion is the unique element whereby gravity is described \cite{Pereira}. The teleparallel connection, which is the most general linear affine connection that is both curvature-less and satisfies the metricity condition, leaves torsion as the base element that describes the gravitational interaction. It is defined by
\begin{equation}\label{Teleparallel-connection}
    \Gamma_{\enspace \mu \nu}^{\sigma} := e_{A}^{\enspace \sigma}\partial_{\mu}e_{\enspace \nu}^{A} + e_{A}^{\enspace \sigma}\omega_{\enspace B\mu}^{A}e_{\enspace \nu}^{B}\,,
\end{equation}
where Greek indices refer to the general manifold while Latin ones refer to its local Minkowski space. The theory is 
local Lorentz invariant due to the presence of an arbitrary spin connection \cite{Cai:2015emx} $\omega_{\enspace B\mu}^{A}$. Hence, for any choice of tetrad $e_{A}^{\enspace \sigma}$, the spin connection balances this freedom with different inertial contributions. This is thus a flat connection and can vanish for particular tetrad choices \cite{Krssak:2015oua} in what is called the Weitzenböck gauge.
These tetrads will be related to other frames by appropriate Lorentz matrices. In GR, the associated spin connections are not inertial and are mainly hidden in the inertial structure of the theory. Both, the tetrad and spin connection are the fundamental dynamical objects on which TG is based, which is invariant under local Lorentz transformations and generally covariant.

There also exists the so-called good tetrads which allow vanishing spin connection components \cite{Toporensky:2019mlg}, and are useful for simplifying the system of equations to solve. This appears when vanishing spin connection and tetrad solutions identically satisfy the spin connection (antisymmetric) field equations. The good tetrads select the class of local Lorentz observers who use the metric $g_{\mu\nu}$ to measure physical distances \cite{Tamanini:2013xya} and do not imply restrictions on the functional form of the Lagrangian contemplated \cite{Tamanini:2012hg}, which as we are going to see can be a function of the torsion scalar ($f(T)$ gravity). The Weitzenböck gauge must be verified 
at a perturbative level to test if it is dually satisfied for the perturbative terms. Tetrad and spin connection pairs produce generally covariant theories which means that field equations will be dynamically equivalent; in other words, any pairs of tetrad and spin connection related through adequate local Lorentz transformations produce the same equations of motion. 

In TG the metric tensor is replaced as the fundamental dynamical variable and supplanted by orthonormal tetrads components \cite{Bamba:2013jqa} that observe the following consistency relations
\begin{equation}
    e^{A}_{\enskip \mu}e_{B}^{\enskip \mu} = \delta^{A}_{B}, \quad e^{A}_{\enskip \mu}e_{A}^{\enskip \nu} = \delta^{\nu}_{\mu}\,,
\end{equation}
and this creates the conditions on which to build inverses of the tetrad fields. One direct application of the transformation rules is that between the Minkowski and general spaces which takes the form
\begin{equation} \label{metric-tetrad}
    g_{\mu\nu} = e_{\enspace \mu}^{A}e_{\enspace \nu}^{B}\eta_{AB}, \quad \eta_{AB} = e_{A}^{\enspace \mu}e_{B}^{\enspace \nu}g_{\mu \nu}\,,
\end{equation}
where the tetrad is the principal variable of these relations. The metric signature to be considered is $(+,-,-,-)$.

It is important to point out that the curvature measured by the Riemann tensor will always vanish in TG due to the curvature-less (flatness) and metric compatibility properties of the teleparallel connection \cite{BeltranJimenez:2019esp}. Thus, we necessitate a different measure of geometric deformation which in TG takes the form of the torsion tensor that is defined as an anti-symmetric property on the connection indices
\begin{equation}\label{torsiontensor}
    T_{\enspace \mu\nu}^{\sigma}:=  2\Gamma_{\enspace [\nu \mu]}^{\sigma}\,.
\end{equation}

This is a measure of the field strength of gravitation (square brackets represent the anti-symmetric operator $A_{[\mu\nu]}=\frac{1}{2}(A_{\mu\nu}-A_{\nu\mu})$) because the non-vanishing torsion is directly related with the gauge field strength. This transforms covariantly under both diffeomorphisms and local Lorentz transformations (as was already mentioned). Just as with the Riemann tensor, the torsion tensor measures the deformation of spacetime.

Other critical tensors can also be defined, one of which is the contortion tensor which measures the difference between the teleparallel and Levi-Civita connections
\begin{equation}\label{contortiontensor}
    K_{\enspace \mu \nu}^{\sigma}:= \Gamma_{\enspace \mu \nu}^{\sigma} - \mathring{\Gamma}_{\enspace \mu \nu}^{\sigma} = \frac{1}{2}\Big(T_{\mu \enspace \nu}^{\enspace \sigma}+T_{\nu \enspace \mu}^{\enspace \sigma}-T_{\enspace \mu \nu}^{\sigma}\Big)\,,
\end{equation}
where $\mathring{\Gamma}{}^{\sigma}{}_{\mu\nu}$ is the Levi-Civita connection\footnote{In this paper all quantities denoted with an over-circle are calculated with the Levi-Civita connection.}. This quantity plays an important role in relating TG with Levi-Civita-based theories.

Another important ingredient in TG is superpotential which is defined as  
\begin{equation}
    S_{A}^{\enspace \mu \nu} := \Big(K_{\quad A}^{\mu \nu}-e_{A}^{\enspace \nu}T_{\quad \alpha}^{\alpha \mu}+e_{A}^{\enspace \mu}T_{\quad \alpha}^{\alpha \nu}\Big)\,.
\end{equation}
The contraction of the superpotential and torsion tensors leads to the torsion scalar, namely
\begin{equation}\label{TorsionScalar}
    T := S_{A}^{\enspace \mu \nu}T_{\enspace \mu \nu}^{A}\,,
\end{equation}
which is solely determined by the teleparallel connection (and consequently by the tetrad and the spin connection) in the same way that the Ricci scalar is dependent only on the Levi-Civita connection. Once again, we have here a representation with a Latin index in the superpotential and the torsion tensor. It turns out that the torsion and Ricci scalars are equal up to a boundary term, i.e. \cite{sebas6}
\begin{equation}\label{scalars}
    \mathring{R} = -T + \frac{2}{e}\partial_{\mu}\Big(eT_{\sigma}^{\enspace \sigma \mu}\Big)+R := -T + B\,,
\end{equation}
where $R$ is the Ricci scalar as calculated with the teleparallel connection which vanishes, $e$ is the determinant of the tetrad field, $e = \det (e^{a}_{\enskip \mu}) = \sqrt{-g}$, and the boundary term, that is a divergence of the torsion vector, is 
\begin{equation}\label{boundaryterm}
    B = \frac{2}{e}\partial_{\mu}\Big(eT^{\mu}\Big)\,.
\end{equation}
Here, the torsion vector is the result of a contraction of two indices in the torsion tensor

\begin{equation}
    T^{\mu} = T_{\sigma}^{\enspace \sigma \mu}\,.
\end{equation}
The TEGR action is expressed as
\begin{equation}\label{TGaction}
    \mathcal{S} = \int d^4x \bigg[\frac{1}{2\kappa^2}T + \mathcal{L}_{\rm m}\bigg]e\,,
\end{equation}
whose dynamical fields are the tetrad field and the spin connection, and where $\kappa^2 = 8\pi G$\footnote{We will be using natural units where $c = 1$.} and $\mathcal{L}_{\rm m}$ represents the Lagrangian for the matter. The variation of such action concerning the tetrad field results in completely equivalent dynamical equations to GR.

The variation of (\ref{TGaction}) concerning the tetrad field produces 
\begin{equation}\label{TEGRfieldeq}
    \dut{E}{A}{\mu} := \frac{1}{e}\partial_{\nu}(eS_{A}^{\enskip \mu \nu})-T_{\enskip \nu A}^{B}S_{B}^{\enskip \nu \mu}+\omega^{B}_{\enskip A\nu}S_{B}^{\enskip \nu \mu}+\frac{1}{2}Te_{A}^{\enskip \mu}=\kappa^{2}\mathcal{T}_{A}^{\enskip \mu}\,,
\end{equation}
where $\mathcal{T}_{A}^{\enskip \mu}$ is the regular energy-momentum tensor of matter. The spin connection field equations turn out to be the anti-symmetric version of these equations. Due to the symmetry of the energy-momentum tensor, this turns out to vanish $E_{[\mu\nu]} = 0$. It is these extra equations that we must also satisfy for our choice of spin connection.

In what follows we are going to see how Eq.~\eqref{TGaction} and Eq.~\eqref{TEGRfieldeq} change when we extend the Lagrangian to an arbitrary function of the torsion scalar.


\subsection{\texorpdfstring{$f(T)$}{} gravity}
\label{subsec:f(T)-background}

In the same way that GR can be generalized, TG can also be extended in different forms \cite{Krssak:2018ywd}. The action in Eq.~(\ref{TGaction}) can be generalized to arbitrary functions of the torsion scalar to produce $f(T)$ gravity \cite{Ferraro:2006jd, Ferraro:2008ey, Bengochea:2008gz, Linder:2010py, Chen:2010va,Farrugia:2016pjh} which follows the same reasoning as $f(\mathring{R})$ gravity. $f(T)$ is the first and more direct extension of TG. However, unlike $f(\mathring{R})$ gravity \cite{Capozziello:2011et, Sotiriou:2008rp, Faraoni:2008mf} that is a fourth order theory under the metric formalism, $f(T)$ gravity produces generally second-order field equations in the tetrad fields. This implies that Lovelock's theorem is, somehow, weakened in TG \cite{Lovelock:1971yv, Gonzalez:2015sha, Bahamonde:2019shr}, and consequently has had interesting effects for constructing scalar-tensor theories of gravity \cite{Bahamonde:2019shr, Bahamonde:2016jqq, Hohmann:2018rwf, Hohmann:2018vle, Skugoreva:2014ena, Hohmann:2018ijr}. 

Some works analyze different $f(T)$ models with some good results; in \cite{Anagnostopoulos:2019miu} some $f(T)$ models are studied in the Bayesian framework, considering the background and the perturbative behaviour simultaneously. It was reported that those specific models, and one of them in particular, had a good effectivity in predicting cosmological data. Another example of the relevance of $f(T)$ models is the one reported in \cite{Bamba:2010wb}; with regards to the topic of dark energy, an $f(T)$ model (specifically, a combined $f(T)$ theory with both logarithmic and exponential terms) was able to allow the crossing of the line that divides the phantom and non-phantom region in the cosmological evolution. This is an important result in the cosmic expansion context. 

The $f(T)$ teleparallel gravity is one of the most studied and used extensions of TG to describe the gravitational interaction. The action to be considered in this case is
\begin{equation}
    \mathcal{S} = \int d^4x \bigg[\dfrac{1}{2\kappa^2}f(T) + \mathcal{L}_{\rm m}\bigg]e\,,
\end{equation}
and its variation concerning the tetrad field leads to
\begin{equation}\label{fTfieldeq}
    -\frac{1}{e}f_{T}\partial_{\nu}(eS_{A}^{\enskip \mu \nu})-S_{A}^{\enskip \mu \nu}\partial_{\nu}f_{T}+f_{T}T_{\enskip \nu A}^{B}S_{B}^{\enskip \nu \mu}-f_{T}\omega^{B}_{\enskip A\nu}S_{B}^{\enskip \nu \mu}-\frac{1}{2}fe_{A}^{\enskip \mu}=\kappa^{2}\mathcal{T}_{A}^{\enskip \mu}\,.
\end{equation}
Notice that when $f(T) = T$, which is the TEGR case, Eq.~(\ref{fTfieldeq}) is reduced to Eq.~(\ref{TEGRfieldeq}), as expected. This extension has several other similarities with GR such as exhibiting the same Gravitational Waves (GW) polarizations \cite{Nunes:2018evm, Farrugia:2018gyz}.

Considering an FLRW metric and an energy-momentum tensor associated with a perfect fluid, from Eq.~(\ref{fTfieldeq}), we obtain the modified Friedmann equations in this case, which are
\begin{equation}\label{HT1}
    H^{2}=\frac{\kappa^{2}}{3}\rho+\frac{1}{6}\Big(T-f+2Tf_{T}\Big)
\end{equation}
and
\begin{equation}\label{HT2}
    \dot{H}=-\frac{\kappa^{2}}{2}(\rho+p)+\dot{H}\Big(1+f_{T}+2Tf_{TT}\Big),
\end{equation}
where $H$ is the Hubble parameter\footnote{The Hubble parameter in cosmic time is defined as $H = \dot{a}(t)/a(t)$, and the dots refer to derivatives concerning the cosmic time.}, and $\rho$ and $p$ are the density and pressure, respectively, associated with the perfect fluid, which has a contribution of matter and radiation; that is, $\rho = \rho_{m} + \rho_{r}$, where $\rho_m$ is the baryonic matter (including dark matter) and $\rho_{r}$ is radiation (including neutrinos). Evaluating Eq.~\eqref{TorsionScalar} with the unperturbed FLRW metric, we can find that \cite{Cai:2011tc}
\begin{equation}
    T = -6H^{2}.
\end{equation}

\noindent The perturbation procedure we will follow is directly on the tetrad field and, consequently, on the rest of the quantities already discussed. The scalar potentials will modify the field equations, and this treatment has to be consistent with the GR perturbed field equations when we take the TEGR case. In the next section, we discuss this process.


\section{\texorpdfstring{$f(T)$}{} scalar perturbations and its standard numerical implementation}
\label{sec:perturbfT}

As it is usually done, we assume that all departures from homogeneity and isotropy in cosmic history have been small. This allows to treat inhomogeneity and anisotropy as first-order perturbations, so that the metric tensor can be expressed up to linear order in the following way
\begin{equation}\label{back+pert-metric}
    g_{\mu\nu} = \Bar{g}_{\mu\nu} + \delta g_{\mu\nu}\,,
\end{equation}
where $\Bar{g}_{\mu\nu}$ is the background value of the metric (which in our case is the flat Friedman-Lemaitre-Robertson-Walker, FLRW metric) and $\delta g_{\mu\nu}$ are the inhomogeneous and anisotropic corrections; these small perturbations satisfy $|\delta g_{\mu\nu}| << 1$. The perturbations can be decomposed into scalars, divergenceless vectors and divergenceless traceless symmetric tensors; these are not coupled to each other by the field equations or conservation equations, so it enables to simplify the results \cite{weinberg}. In conformal time\footnote{We are considering $\mathcal{H}=a'(\tau)/a(\tau)$, where the Hubble parameter is written in conformal time and primes denote derivatives concerning the proper time.}, the most general perturbed metric can be written as
\begin{equation}\label{Pert-general-metric}
    \delta g_{\mu\nu}= \begin{bmatrix}
        a^{2}(2\phi) & a(\partial_{i}\mathcal{B} + \mathcal{B}_{i})\\
        a(\partial_{i}\mathcal{B} + \mathcal{B}_{i}) & 2a^{2}(\psi\delta_{ij} + \partial_{i}\partial_{j}h + 2\partial_{(i}h_{j)} + \frac{1}{2}h_{ij})
    \end{bmatrix}\,.
\end{equation}
Notice that we have four scalars $\lbrace\phi, \mathcal{B}, \psi, h\rbrace$, 2 vectors $\lbrace\mathcal{B}_{i}, h_{j}\rbrace$ and 1
one tensor $\lbrace h_{ij}\rbrace$. The vector perturbations decay rapidly in their propagation, so they have not played a big role in cosmology, while the tensor perturbations are associated with gravitational waves. These space-time ripples could be directly measured by LIGO \cite{LIGO, Krolak:2017adi}, 
however, since we are interested in matter fluctuations and structure formation we will focus on scalar perturbations only. To keep those components and get rid of the rest (due to the decoupling between them) we must make a proper selection 
of the quantities involved due to the gauge freedom. The Newtonian gauge (or longitudinal gauge) is the one used for this type of perturbation since it restricts the metric to be applicable only for the scalar mode (the vector and tensor degrees of freedom are eliminated). With this at hand, we take $\mathcal{B} = 0$ and $h = 0$, so the line element of the general metric tensor \eqref{back+pert-metric} looks like
\begin{equation}\label{Pmetric}
    ds^{2}=a^{2}(1+2\phi)d\tau^{2}-a^{2}(1-2\psi)dx_{i}dx^{i}\,,
\end{equation}
where $\psi$ and $\phi$ are the only scalar potentials. These are called the so-called Bardeen potentials, and they will give the required dynamics. The main goal is to find an expression for these scalars through the perturbed field equations and use those expressions to plot the CMB power spectra. The first step is to remember that in TG we naturally work with the tetrad formalism. Analogously to \eqref{back+pert-metric}, the tetrad field can be written as
\begin{equation}\label{back+pert-tetrad}
    e^{A}_{\enskip\mu} = \Bar{e}^{A}_{\enskip\mu} + \delta e^{A}_{\enskip\mu}\,,
\end{equation}
where $\Bar{e}^{A}_{\enskip\mu}$ is the background tetrad (related to the flat FLRW metric), and where $\delta e^{A}_{\enskip\mu}$ is the perturbation term.

Using Eq.~\eqref{metric-tetrad}, we can deduce that the most general tetrad that recovers the metric in Eq.~\eqref{Pmetric},
\begin{equation}\label{Ptetrad}
    e^{A}_{\enskip\mu}= \begin{pmatrix}
        a(1+\phi) & \partial_{x}\beta & \partial_{y}\beta & \partial_{z}\beta\\
        a\partial_{x}\beta & a(1-\psi) & a\partial_{z}\sigma & -a\partial_{y}\sigma\\
        a\partial_{y}\beta & -a\partial_{z}\sigma & a(1-\psi) & a\partial_{x}\sigma\\
        a\partial_{z}\beta & a\partial_{y}\sigma & -a\partial_{x}\sigma & a(1-\psi)
    \end{pmatrix}\,,
\end{equation}
where we notice that two degrees of freedom do not contribute to the metric; i.e. $\beta$ and $\sigma$. We can call these two scalars gauge degrees of freedom since they do not change the background field equations. Nevertheless, they appear at the perturbative level. The generally perturbed tetrad \eqref{Ptetrad} is Weitzenb\"{o}ck gauge compatible and, therefore, the corresponding spin connection vanishes\footnote{At perturbative level, the spin connection has vanished at background level, moreover it must reappear in the perturbed field equations. However, this is not possible because the scalar modes are gravitational effects, and the spin connection is purely inertial. Therefore, the components of the spin connection are zero at all levels if the Weitzenböck gauge is correctly applied.}.

The perturbed components of the corresponding energy-momentum tensor are:
\begin{equation}\label{Ene-Mo00}
    \delta T_{00} = -\delta\rho\,,
\end{equation}
\begin{equation}\label{Ene-Mo0i}
    \delta T_{0}^{\enskip i} = -\frac{1}{a^2}(\rho+p)\partial^{i}v\,,
\end{equation}
\begin{equation}\label{Ene-Moi0}
    \delta T_{i}^{\enskip 0} = (\rho+p)\partial_{i}v\,,
\end{equation}
\begin{equation}\label{Ene-Moij}
    \delta T_{ij} = \delta_{ij}\delta p+\partial_{i}\partial_{j}\Pi^{S}\,,
\end{equation}
where $\Pi^{S}$ is the scalar part of the anisotropic stress and $v$ is the velocity potential. These two quantities represent dissipative corrections to the energy-momentum tensor from the perfect fluid. Introducing the perturbed quantities into \eqref{fTfieldeq}
we arrived to the perturbed $f(T)$ gravity field equations, just as \cite{Zheng:2010am}. Then, the resulting equations are
\begin{equation}\label{FinalTG00}
    \frac{1}{2}\kappa^{2}\delta\rho = \frac{f_{T}}{a^2}\partial^{2}\psi-12\frac{\mathcal{H}^{2}}{a^4}f_{TT}\partial^{2}\zeta-3\frac{\mathcal{H}}{a^2}\Big(f_{T}-12\frac{\mathcal{H}^{2}}{a^2}f_{TT}\Big)\Big(\psi'+\mathcal{H}\phi\Big)\,,
\end{equation}
\begin{equation}\label{FinalTGi0}
    \frac{1}{2}\kappa^{2}(\rho+p)\partial^{i}v = -\frac{f_{T}}{a^2}\partial^{i}\Big(\psi'+\mathcal{H}\phi\Big)+12\frac{\mathcal{H}}{a^4}f_{TT}\Big(\mathcal{H}'-\mathcal{H}^{2}\Big)\partial^{i}\psi\,,
\end{equation}
\begin{equation}\label{FinalTG0i}
    \frac{1}{2}\kappa^{2}(\rho+p)\partial_{i}v = -\frac{1}{a^2}\Big(f_{T}-12\frac{\mathcal{H}^2}{a^2}f_{TT}\Big)\partial_{i}\Big(\psi'+\mathcal{H}\phi\Big)-4\frac{\mathcal{H}}{a^4}f_{TT}\partial_{i}\partial^{2}\zeta\,,
\end{equation}
\begin{equation}\label{FinalTGij}
    \frac{1}{2}\kappa^{2}a^{2}\partial_{i}\partial^{j}\Pi^{S} = f_{T}\partial_{i}\partial^{j}(\phi-\psi)+12\frac{f_{TT}}{a^2}\Big(\mathcal{H}'-\mathcal{H}^{2}\Big)\partial_{i}\partial^{j}\zeta\,, \quad \text{when}\quad i\neq j
\end{equation}
\begin{equation*}
    \frac{1}{2}\kappa^{2}\delta p = \frac{f_{T}}{a^2}\bigg[\psi''+2\mathcal{H}\psi'+\mathcal{H}\phi'+2\Big(\mathcal{H}'-\mathcal{H}^{2}\Big)\phi+3\mathcal{H}^{2}\phi-\frac{1}{2}\partial^{2}(\psi-\phi)\bigg]+\frac{f_{TT}}{a^4}\bigg[-12\mathcal{H}^{2}\psi''-36\mathcal{H}\mathcal{H}'\psi'
\end{equation*}
\begin{equation*}   
    -12\mathcal{H}^{3}\phi'-\Big[60\Big(\mathcal{H}'-\mathcal{H}^{2}\Big)\mathcal{H}^{2}+36\mathcal{H}^{4}\Big]\phi+\Big[8\Big(\mathcal{H}'-\mathcal{H}^{2}\Big)+4\mathcal{H}^{2}\Big]\partial^{2}\zeta+4\mathcal{H}\partial^{2}\zeta'\bigg]
\end{equation*}
\begin{equation}\label{FinalTGii} 
    +12\frac{f_{TTT}}{a^6}\Big(\mathcal{H}'-\mathcal{H}^{2}\Big)\mathcal{H}^{2}\bigg[12\mathcal{H}(\psi'+\mathcal{H}\phi)-4\partial^{2}\zeta\bigg], \quad \text{when}\quad i=j\,.
\end{equation}
In these latter equations, $\zeta = \mathcal{H}\beta$, and to get correspondence with GR the constant $\sigma$ has been considered to be zero, whereas $\beta$ is still arbitrary. This parameter is always set to zero too, but we could assume it is a degree of freedom and constrain it. 

Up to this point, the equations are valid for a general $f(T)$ Lagrangian and express the equations of motion of a perturbed energy-momentum fluid assumed to be the universe at early times. In \cite{Farrugia:2016pjh}, a particular generalization of (\ref{FinalTG00})-(\ref{FinalTGii}) was made.

Considering the TEGR limit we then obtain
\begin{equation}
    \kappa^{2}\delta\rho = -6\frac{\mathcal{H}}{a^2}(\psi'+\mathcal{H}\phi)+2\frac{\partial^2}{a^2}\psi\,,
\end{equation}
\begin{equation}
    \kappa^{2}(\rho+p)\partial^{i}v = -\frac{2}{a^2}\partial^{i}\Big(\psi'+\mathcal{H}\phi\Big)\,,
\end{equation}
\begin{equation}
    \kappa^{2}(\rho+p)\partial_{i}v  = -\frac{2}{a^2}\partial_{i}\Big(\psi'+\mathcal{H}\phi\Big)\,,
\end{equation}
\begin{equation}
    \phi=\psi\,, \quad \text{when}\quad i\neq j
\end{equation}
\begin{equation}
    \kappa^{2}\delta p = \frac{2}{a^2}\Big[(\mathcal{H}^{2}+2\mathcal{H}')\phi+\mathcal{H}\phi'+\psi''+2\mathcal{H}\psi'\Big]+\frac{\partial^2}{a^2}(\phi-\psi)\,, \quad \text{when}\quad i=j
\end{equation}
which are the GR perturbed field equations already known in the literature (see, for instance, Ref.~\cite{Amendola}).

At this point, it is important to describe the possible approaches to proceed in the cosmological constraint analysis for the theory. On one hand, we have the so-called
\begin{itemize}
\item \textbf{Fluid-like approach (FL)}. This method assumes a standard primordial perturbation pattern for a perfect fluid which is named \textit{torsion fluid}, which will be responsible for the current cosmic expansion. One full description can be found in \cite{Ferraro:2012wp,Nunes:2016plz,Nunes:2018evm,Nunes:2018xbm,Benetti:2020hxp},
however, here we will discuss some generalities. From the TEGR modified Friedmann equations \eqref{HT1} and \eqref{HT2}, we can put together the extra torsion terms in each equation as an effective energy density $\rho_{T}$ and pressure $p_{T}$. These two quantities contain now the modifications related to the torsion fluid, which can be seen as the effective standard density and pressure for a dark energy model. Furthermore, we can define an Equation-of-State (EoS) with both quantities through a standard barotropic relation $p=w\rho$. Analogously, we can write $w_{T}=p_{T}/\rho_{T}$, for the torsion fluid. This relation is called the \textit{effective torsion Equation-of-State}, where $w_{T}$ is the effective state parameter. Therefore, the evolution of the total density can be written as $E(a)=H(a)/H_{0}$, where the effective energy density is included through the Eqs.~(\ref{HT1})-(\ref{HT2}). Under this approach, we can employ the Boltzmann equations using the effective torsion and analyse the theoretical observational predictions for the CMB anisotropies, including the components for the CMB angular power spectrum. As it was shown in \cite{Benetti:2020hxp}, this approach seems to relax the $H_{0}$ tension in a better way than other $f(T)$ models but worsens the $\sigma_{8}$ tension. This last effect can be because using this fluid-like approach as the effective torsion EoS will have a strong correlation between $H_0$ and $\sigma_8$.

\end{itemize}
On the other hand, the approach we propose in this work is the
\begin{itemize}
\item \textbf{Non fluid-like approach (NFL)}. In this case, Eqs.~\eqref{HT1} and \eqref{HT2}, which come from a standard FLRW metric, are no longer valid at the perturbative level, and so we need to perturb the tetrad field (in the Newtonian gauge) just as in \eqref{Ptetrad}. Also, we need to consider the perturbed components of the energy-momentum tensor for a perfect fluid, which are given by Eqs.~\eqref{Ene-Mo00}-\eqref{Ene-Moij}. The final field equations are, consequently, rather different from the first approach; these are shown in \eqref{FinalTG00}-\eqref{FinalTGii}. The next step will be to choose the desired $f(T)$ model (in our case the power-like model (see Sec.\ref{subsec:fT_pert}) and substitute the function $f(T)$ (and its derivatives $f_T, f_{TT}$) into the perturbed field equations. Afterwards, it is possible to study the evolution of the scalar perturbations (see Sec.\ref{sec:perturbfT}) through the different components of the CMB angular power spectrum. As we can notice, the two approaches are different in form and essence; the fluid-like approach does not perform the perturbations in the modified field equations, since just assumes a new type of gravitational component created by torsion as an effective fluid, and so the contribution to the matter fluctuations is hidden in the structure of the energy density. Instead, the non-fluid-like performs a direct perturbation in the tetrad field, leading to the perturbed field equations for the $f(T)$ gravity. In this latter case, the effect over the CMB anisotropies is given directly and explicitly, which in principle leads to deviations from the effective torsion fluid.

\end{itemize}


\subsection{Perturbed $f(T)$ power law cosmology: numerical code adaptation expressions}
\label{subsec:fT_pert}

For our perturbation analysis, we consider a specific $f(T)$ model and derive the perturbed field equations for this model to obtain the power spectra, respectively. We are going to choose a power law model, which has the following  mathematical form
\begin{equation}\label{PLM}
    f(T) = T + \alpha (-T)^{b}\,,
\end{equation}
where $\alpha$ and $b$ are the constant model parameters. This model is a good candidate since it produces the same effect as dark energy and exhibits phases dominated by matter and radiation \cite{Bengochea:2008gz}. In addition, it does not exclude the $\Lambda$CDM paradigm and is in excellent agreement with current experimental bounds \cite{LeviSaid:2020mbb}. 

Considering Eq.~\eqref{HT1} at present, it is possible to find that
\begin{equation}\label{alpha-parameter}
    \alpha = (6\mathcal{H}_{0}^{2})^{1-b}\Big(\frac{1-\Omega_{m0}-\Omega_{r0}}{1-2b}\Big)\,,
\end{equation}
where $\mathcal{H}_{0}$ is the Hubble parameter measured at present time, $\Omega_{m0}$ is the current density parameter associated with the sum of baryonic and dark matter and $\Omega_{r0}$ is the current density parameter associated with the sum of radiation and neutrinos. Consequently, we can see that $b$ is the only real free parameter by initial conditions.

The two perturbed field equations using the power law model Eq.~\eqref{PLM}, are
\begin{equation}\label{perturbpsi}
    \psi = \phi - 12\bigg[\frac{\alpha b(b-1)(6\mathcal{H}^{2})^{b-2}}{1-\alpha b(6\mathcal{H}^{2})^{b-1}}\bigg]\Big(\frac{a''}{a^3}-2\mathcal{H}^{2}\Big)a\mathcal{H}\beta-\frac{9}{2}\Big(\frac{a^2}{k^2}\Big)\bigg[\frac{(\rho+p)\Upsilon}{1-\alpha b(6\mathcal{H}^{2})^{b-1}}\bigg]\,,
\end{equation}
and 
\begin{equation}\label{perturbphiprime}
    \phi' = -a\mathcal{H}\psi+\frac{4a^{2}\mathcal{H}^{2}\alpha b(b-1)(6\mathcal{H}^{2})^{b-2}\beta}{1-\alpha b(6\mathcal{H}^{2})^{b-1}-12\mathcal{H}^{2}\alpha b(b-1)(6\mathcal{H}^{2})^{b-2}} + \frac{1.5(\frac{a^2}{k^2})(\rho+p)\theta}{1-\alpha b(6\mathcal{H}^{2})^{b-1}-12\mathcal{H}^{2}\alpha b(b-1)(6\mathcal{H}^{2})^{b-2}}\,,
\end{equation}
where we have made $\psi \leftrightarrows \phi$. This last step was done because from the beginning we chose $\phi$ to be in the temporal component of the metric, and $\psi$ in the spatial component, as we can see from \eqref{Pmetric}. However, some authors prefer the inverse notation ($\phi$ in the spatial component and $\psi$ in the temporal), and in fact, the Boltzmann code that we use will adopt this last convention, so to agree with the interchange between the two scalars was performed.
The expressions for $\psi$ and $\phi'$ found in \eqref{perturbpsi} and \eqref{perturbphiprime} come from the perturbed field equations when the power law model is substituted. In papers \cite{Chen:2010va, Li:2011wu, Zheng:2010am} the scalar perturbations in $f(T)$ gravity were studied, but \eqref{perturbpsi} and \eqref{perturbphiprime} were not derived. In \cite{Nunes:2018xbm} they used the same Boltzmann code we are interested in for the same purposes but did not give an expression for these scalars. Moreover, we have written the expressions for the two scalars with the gauge degree of freedom $\beta$. The Boltzmann code we will use has particular conventions; units assume $c=1$, times and distances are in Mpc, and all densities and pressures in the code are some rescaled
variables (in $\text{Mpc}^2$):

\begin{equation}\label{class-density-pressure}
    \rho = \frac{8\pi G}{3}\rho_{\text{physical}}, \quad p = \frac{8\pi G}{3}p_{\text{physical}},
\end{equation}
where \textit{physical} denotes that the quantity has its usual cosmological value, consequently, \eqref{class-density-pressure} are the density and pressure which were introduced in \eqref{perturbpsi} and \eqref{perturbphiprime}. 

In the expressions for the Bardeen potentials, $\Upsilon$ is the shear, $\theta$ is the velocity divergence and $k$ is the wave number that appears when moving to Fourier space. Equations \eqref{perturbpsi} and \eqref{perturbphiprime} represent the Bardeen potentials which describe the matter perturbations. These potentials are gauge invariant. The linear treatment is often chosen because it has been seen that the first-order theoretical approximation is more than enough to adjust the inhomogeneity and anisotropy measured from the CMB. In the next section, we explain how the code works and the changes that must be implemented to define a non-fluid-like approach using this model.

\section{Boltzmann code architecture and results for the \texorpdfstring{$f(T)$}{} power law model} 
\label{sec:BC}

To proceed with the non-fluid approach described in this work, we require a set of implementations in the Boltzmann code architecture. A Boltzmann code refers to numerical codes that include the full Boltzmann equations, e.g. photons, neutrinos, baryons, and cold dark matter, in addition to the equations of motion. 
The physics extracted from these codes are related to the CMB power spectra spectrum that includes relevant information on early and late times dynamics of the Universe. The Boltzmann code written in language C \cite{NumRecinC} most used for the precision cosmology community is the so-called Cosmic Linear Anisotropy Solving System (\texttt{CLASS}) software \cite{class,classii}\footnote{\href{https://lesgourg.github.io/class\_public/class.html}{lesgourg.github.io/class\_public/class}}, which can compute several predictions given a specific model(s) for a certain theory of gravity. 

The quantities and equations employed in this software require to be expressed in particular units, e.g. the conformal Hubble parameter $\mathcal{H}$ and the wavenumbers $k$ should be written in 1/Mpc, where $\tau$ has units of Mpc. The CMB multipoles ($C_l$'s) are dimensionless.

\begin{figure}[H]
\includegraphics[width=15.cm]{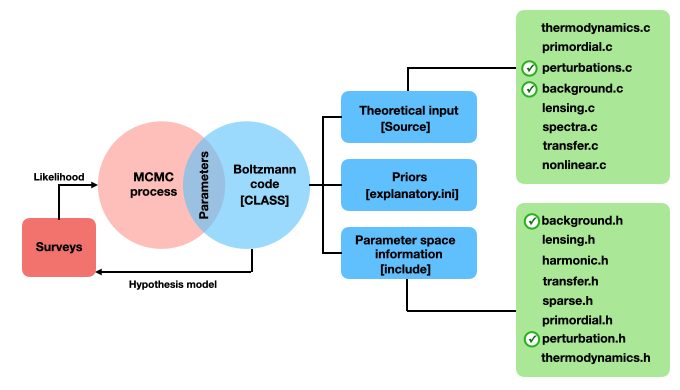}
\centering
\caption{MCMC + Boltzmann code architecture for precision cosmology. In this diagram, we denote the test run for a cosmological parameter space. The numerical process starts in the red colour region. The blue colour region is the main core to be modified for a non-fluid-like approach. The green colour blocks contain the \texttt{CLASS} files in language C and declarations files. According to steps 1-2-3, the files to be modified are indicated with circle tickmarks along the name of the file.}
\label{fig:architectureCLASS}
\end{figure}

Following the standard approach,
to proceed with Boltzmann codes within \texttt{CLASS} scheme we start with perturbations obtained for a specific cosmological model, e.g. $\Lambda$CDM model, and subsequently, we add the equations that include effective-like fluid forms. However, in this work, we carry on a modified version that includes our non-fluid-like approach described in Sec.~\ref{sec:perturbfT}. To have this modified version we are required to proceed with an architecture like the one described in Figure \ref{fig:architectureCLASS}. In this figure, from left to right, first, we require a code from which we can take a sample of the model parameter to predict data. This process is called the Monte Carlo Markov Chain (MCMC). The Boltzmann code (in this architecture, \texttt{CLASS}) will give a set of cosmological parameters that should be compared with the theoretical model using the data stored like surveys. The numerical process continues until we reach a rate of convergence enough to obtain a global best fit for each parameter.

For our non-fluid-like approach, we need to modify some \textit{files} within \texttt{CLASS}. We refer to the files' names in italics. There are three principal cores to consider:

\begin{enumerate}
    \item \textbf{Theoretical input} [\textit{source}]. This part of the architecture includes all the theoretical information related to cosmological dynamics. To implement the non-fluid-like approach we require to modify the current equations by the ones presented in \eqref{fTfieldeq}-\eqref{FinalTG00}-\eqref{FinalTGii}. [\texttt{CLASS} implementation]: in \textit{background.c} file\footnote{The .c denotes language C code.} we rewrite the existing Friedmann equations by our equations \eqref{HT1}-\eqref{HT2} for the $f(T)$ power law model. In \textit{perturbations.c} file we should rewrite the perturbed equations by \eqref{perturbpsi}-\eqref{perturbphiprime}. Additionally, in \textit{input.c} file we must define what kind of variable is $b$ (how the code should read it), and here we can explicitly write the form of $\alpha$ which appears in \eqref{alpha-parameter} (this last step is optional because one can just substitute the explicit expression into the perturbed equations instead of using $\alpha$).
    \item \textbf{Priors} [\textit{explanatory.ini} file]. Contains the numerical intervals of the free cosmological parameters of interest given in the model analyzed. [\texttt{CLASS} implementation]:
     in \textit{explanatory.ini} file we introduced the physical intervals associated with our free parameters indicated in \eqref{Ptetrad}-\eqref{PLM}; that is, the gauge degree of freedom $\beta$ which appears in the generally perturbed tetrad and the unique free parameter of the power law model $b$. The rest of the cosmological parameters remain. Also in this file, it is necessary to define the gauge associated with the perturbations; since we are dealing with scalar perturbations we must choose the Newtonian gauge.
    \item \textbf{Parameter space information} [\textit{include}]. This part of the architecture contains the definition of all the cosmological parameters of interest, therefore, we are required to add here the new parameters related to the $f(T)$ model. [\texttt{CLASS} implementation]: in \textit{background.h} file\footnote{The .h is because these are header files containing declarations.} we added the free parameters of interest associated with our model described by equations \eqref{Ptetrad}-\eqref{PLM}.
\end{enumerate}

\subsection{\texttt{Python} wrapper adaptation}
\label{subsec:Python-wrapper}

It is useful to work with the \texttt{Python} Wrapper of \texttt{CLASS}, which enables to modification of the functionality of the original numerical code in \texttt{CLASS} described at the beginning of this section but without changing its architecture. Moreover, for many users, \texttt{Python} is friendlier than the C language which is the one used in the Boltzmann original code. Under the same steps, the installation details of the Python Wrapper in \texttt{CLASS} are given in \cite{Pywrapper}, however, in this subsection, we will just mention the files which should be modified to implement the non-fluid approach described. 

In the \texttt{CLASS} directory, there is a folder called \texttt{Python}, which contains several files 
Each of these files is meant to fulfil a particular function so the wrapper works properly; for instance, \textit{extract\_errors.py} file is in charge of printing the possible errors (in the system, in how \texttt{Python} runs) that may arise while running the wrapper. In the case of \textit{interface\_generator.py} file, it reads the headers of the files and generates an interface. From a cosmological point of view, this means that would be able to see which physical content is referring to the wrapper in the output. Finally, there will be six files minimum in the \texttt{Python} folder, however, for our purposes, we just need to change the ones related to the definitions of cosmological parameters. This is important since we are introducing a new free parameter $p$ due to the power law model given in Eq.~(\ref{PLM}), and has to be included in these declarations. To adapt our non-fluid-like approach for the model described in Sec.~\ref{subsec:fT_pert}, we need to change the following files:

\begin{enumerate}
    \item \textit{cclassy.pxd}. In this file, we must define the quantities which shall be used in the \texttt{Python} Wrapper as input, output and manipulation. Consequently, the cosmological parameters should be defined in this file; in particular, in a section inside the code called \textit{cdef struct background}. In this file, we should define the free parameter(s) of the cosmological model at hand, which in our case is the parameter $b$ from the $f(T)$ power law model given in Eq.~(\ref{PLM}), therefore it can adopt real values. 
    \item \textit{classy.pyx}. There is a list of elements in this file that \texttt{MontePython} should interpret as cosmological parameters such as $h$, $\Omega_{m}$ and so on. In a section inside the code called \textit{def get\_current\_derived\_parameters(self, names)} these changes should be performed. Once again, the free parameters need to be added so the wrapper can read their names correctly. 
    \item \textit{classy.c}. This has to do with the ``cython'' package, which is the tool that creates the bridge between \texttt{CLASS} and \texttt{Python}. The free parameter $b$ of the $f(T)$ model (and in general, any new free parameter of any cosmological model under consideration) must be included in several parts of the file. 
\end{enumerate}

\noindent The folder \textit{scripts} within \texttt{CLASS} must contain the \texttt{Python} notebook, this is a \textit{.py} file (that you create) in which all the instructions are given: here, you import packages (write the names of the mathematical tools you will need), assign values to the cosmological parameters, and type the commands for the evolution of certain functions of interest, for instance, you can write the instructions to plot the cosmic distances, the different CMB power spectra, etc. Everything in this notebook (file) must be written in \texttt{Python} language.


\subsection{$f(T)$ non-fluid like approach: numerical analysis}
\label{subsec:Analysis}

\noindent To compare the fluid-like and the non-fluid-like approaches, we have considered the following vanilla cosmological parameters:
$\Omega_{b}h^{2} = 0.022161$,
$\Omega_{c}h^{2} = 0.11889$,
$\tau = 0.0952$,
$h = 0.6777$,
$\ln 10^{10}A_{s} = 3.0973$,
$n_{s} = 0.9611$, \cite{Stolzner:2017ged}.
Also, we extended our comparisons using current CMB measurements from the following catalogs:

\begin{itemize}
    \item \textbf{CMB angular power spectrum baseline}. To obtain the TT, EE and TE components, provided by Planck 2018 in \cite{plancklegacy}, we consider 2499 data points for each polarisation component. One of the goals of Planck Collaboration has been to measure the temperature and polarization of the CMB spatial anisotropies, and the current constraints of its data for the standard cosmological model $\Lambda$CDM \cite{Planck:2018nkj}. The full information of the TT, EE and TE components of the CMB power spectra can be found in Ref.\cite{Planck:2019nip}. 
    \item \textbf{SDSS DR7 LRG likelihood}. Obtained from \cite{CC-SDSS}, including 45 points for the matter power spectrum $P(k)$. The Sloan Digital Sky Survey (SDSS) is a kind of catalog which maps one-quarter of the entire sky and computes a redshift survey of astrophysical objects like galaxies, quasars and stars, being the Data Release 7 (DR7) the seventh major catalog which includes images, spectra and their respective redshifts \cite{SDSS-DR7}. In this case, the full information was obtained from Luminous Red Galaxies (LRG) \cite{10.1111/j.1365-2966.2008.13179.x}.
    \item \textbf{ISW data from SDSS catalog}. Provided in \cite{ISW-SDSS}, this catalog consists of a baseline with 100 data points for the galaxy auto-correlation shear-shear angular power spectrum. Furthermore, it contains information on the Integrated Sachs-Wolfe (ISW) effect, which describes how photons adopt a blueshift or redshift while they travel through time-variable potentials between the last scattering epoch and $z=0$. Also, we found in the catalog information of the shear-shear angular power spectrum, which represents the deformation of light due to the scalar perturbations i.e. presence of matter.
    \end{itemize}
Studying the evolution of linear scalar perturbations is important since these affect the anisotropies of the CMB. The different angular power spectra, but especially the TT component, may be affected by the TG. We expect that the scalar perturbations in $f(T)$ gravity can cause significant changes in the anisotropies of the CMB at all angular scales. The fact that the CMB temperature power spectrum changes implies corrections in the structure formation and in the later cosmic expansion, which could relax the $\sigma_{8}$ and the $H_{0}$ tensions. This is the reason why the CMB angular power spectrum datasets are important to see the corresponding deviations due to the $f(T)$ teleparallel power law model proposed. Similar motivations arise for other datasets. For example, the SDSS DR7 LRG baseline will help to see deviations from the matter power spectrum, and therefore to see how the presence of the matter component could change. In the case of the ISW data from the SDSS catalog, this will help us to observe the changes over the shear-shear angular power spectrum and, consequently, to observe how the strength of the gravitational lensing due to the scalar perturbations is weakened/increased.

After performing the code changes described above, we are now ready to run some observational tests using the catalogues described. For this goal, we set the numerical values (in the \textit{explanatory.ini} file) for the cosmological parameters given at the beginning of this subsection, as well as for our free parameter $b$; in our case, we chose $b=0.37$, which is a considerably large value in the power law model to study a significant deviation from the standard $\Lambda$CDM model. In the same file, we set the number of cross-correlation spectra that we wanted to calculate; we chose in the instruction \textit{non\_diagonal=0}, which means only auto-correlation. After that, we can ask for the code for the corresponding output spectra, the temperature, matter and galaxy lensing potential. Finally, we run the code and obtain the desired output files. Using this process, in this work we performed the fluid-like approach and the non-fluid-like approach. Also, to get a comparison with the standard $\Lambda$CDM model we used the non-fluid-like approach with $b=0$, which recovers such a standard model. Consequently, we got three final groups of files which we proceeded to plot along with the datasets for each case: we used the CMB angular power spectrum datasets to obtain the TT, TE and EE components together with the three model predictions. The same with SDSS DR7 LRG likelihood for the matter power spectrum and with ISW data from the SDSS catalog for the first shear-shear auto-correlation angular power spectrum. 

\begin{figure}[H]
\centering
\includegraphics[width=0.32\textwidth]{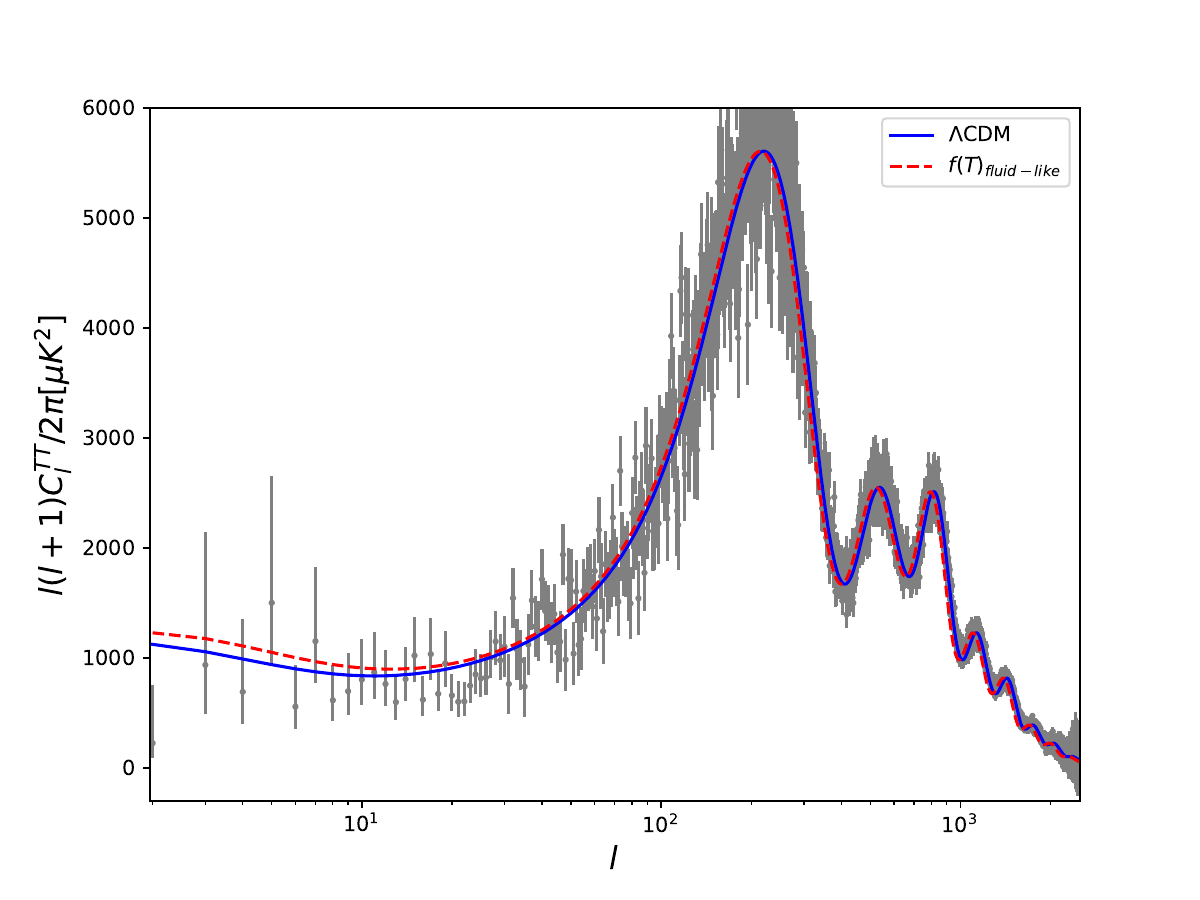}
\includegraphics[width=0.32\textwidth]{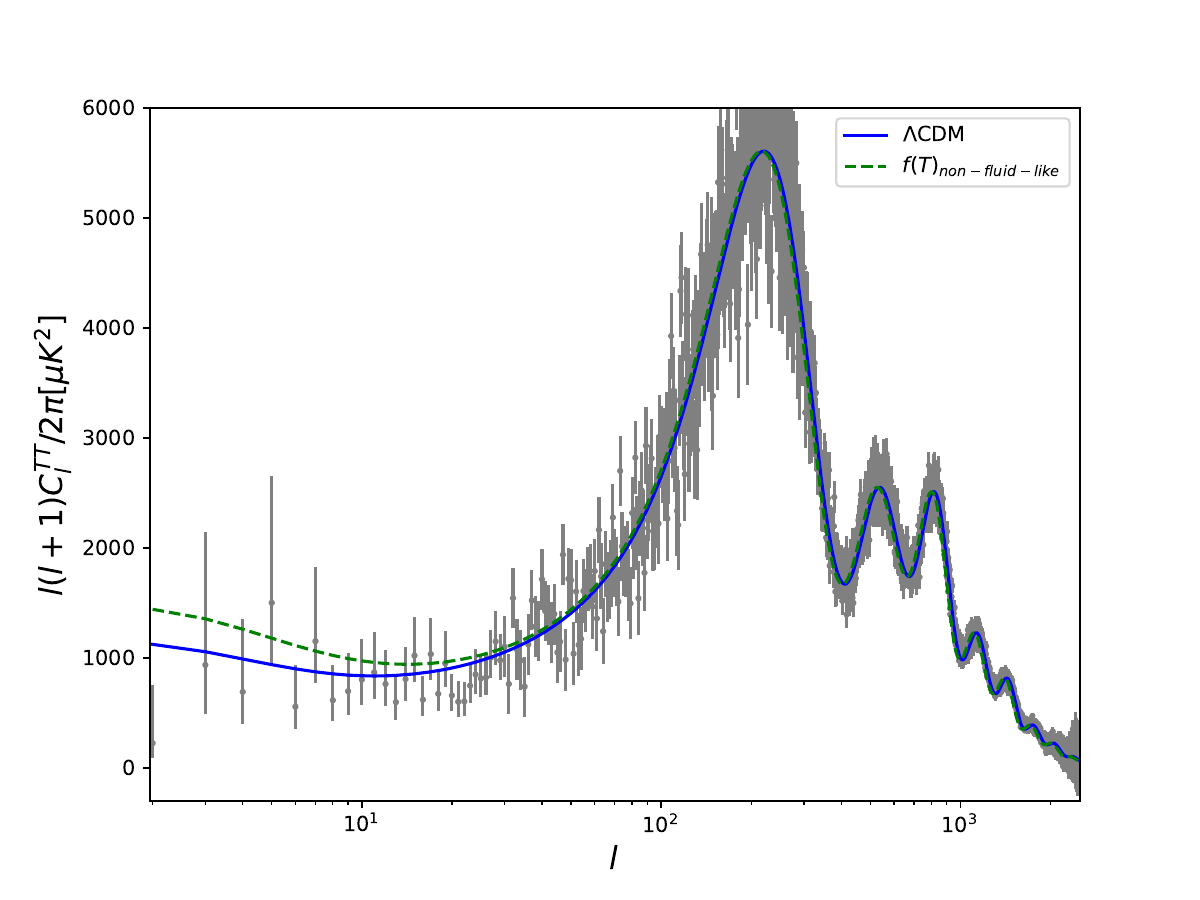}
\includegraphics[width=0.32\textwidth]{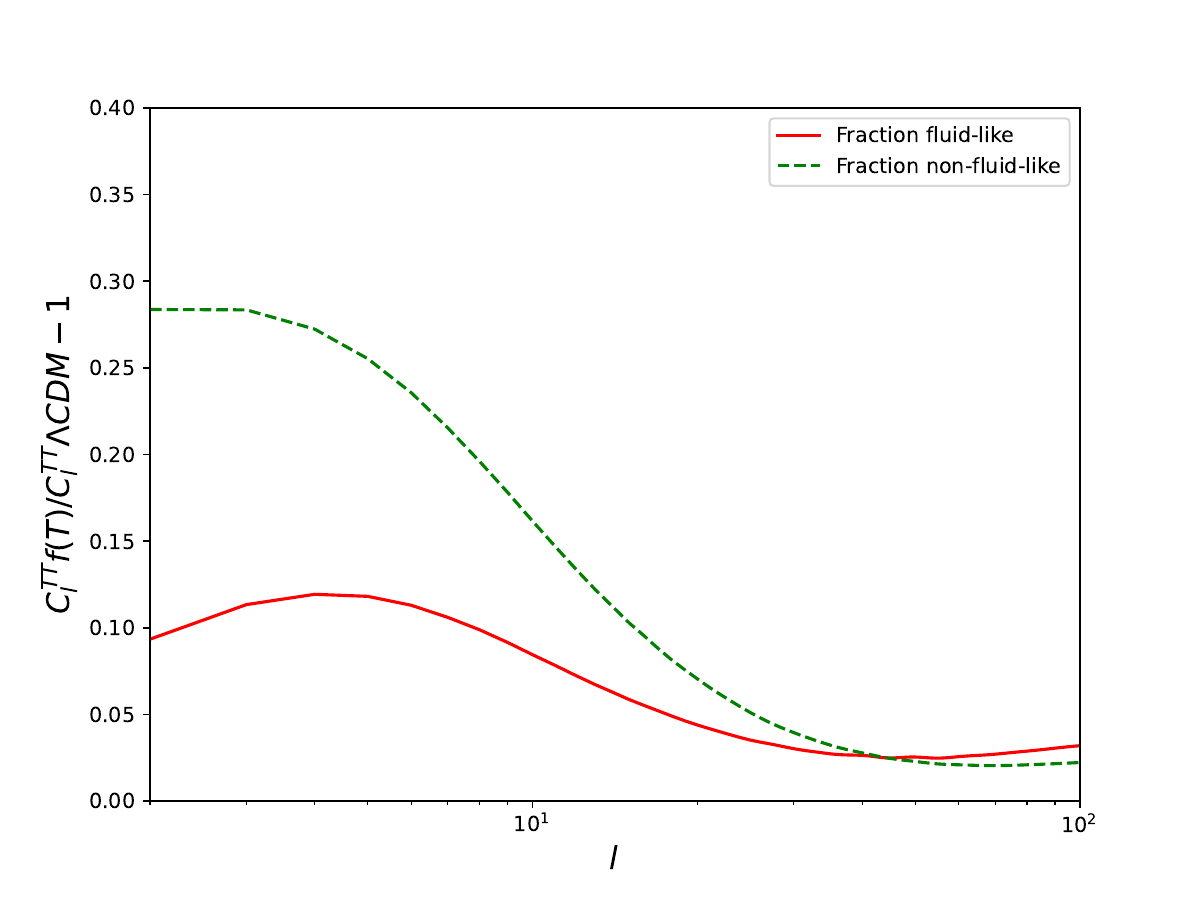}
\caption{Temperature (TT) power spectra for $f(T)$ power law model with $b=0.37$ (see Eq.~\eqref{PLM}) using the fluid-like approach (left) and non-fluid-like approach (middle). The $\Lambda$CDM model is also given (blue curve). In both of these plots, the grey dots correspond to Planck 2018 data \cite{plancklegacy} with their 1$\sigma$ error bars. Furthermore, we show the percentage difference in the temperature (TT) power spectrum between approaches (right). The green curve corresponds to the ratio of the non-fluid-like approach over $\Lambda$CDM, while the red curve is the ratio of the fluid-like approach over $\Lambda$CDM.}
\label{fig:TT}
\end{figure}

\textit{ On the TT power spectrum.-} In Figure \ref{fig:TT} we show the TT power spectrum. In the left figure, we compare the fluid-like approach (red and dashed curve) with $\Lambda$CDM (blue and solid curve). The TT CMB angular power spectrum dataset was also included in the plot (grey points and error bars). In the middle figure, we did it for the non-fluid-like approach (green and dashed curve). We also computed the difference between approaches in the right figure, using the expression $\frac{{C^{TT}_l}_{f(T)\text{approach}}}{{C^{TT}_l}_{[\Lambda\text{CDM}]}}-1$. From these three plots, we can see that the non-fluid-like approach shows a large deviation from $\Lambda$CDM in comparison to the fluid-like approach at low multipoles. This deviation corresponds to a 20\% difference between approaches. Consequently, the non-fluid-like approach causes a larger $C^{TT}_{l}$ variation at early epochs, rather than $\Lambda$CDM and the fluid-like approach. This means that, at least in the early universe, the fluid-like and the non-fluid-like approaches are not equivalent at the perturbative level. Since the non-fluid-like approach consists of the implementation of the `real' perturbed field equations, this implies that the fluid-like scheme loses dynamical information about the universe, at least at low multipoles, and so it should be avoided in this epoch of the universe.

\begin{figure}[H]
\centering
\includegraphics[width=0.32\textwidth]{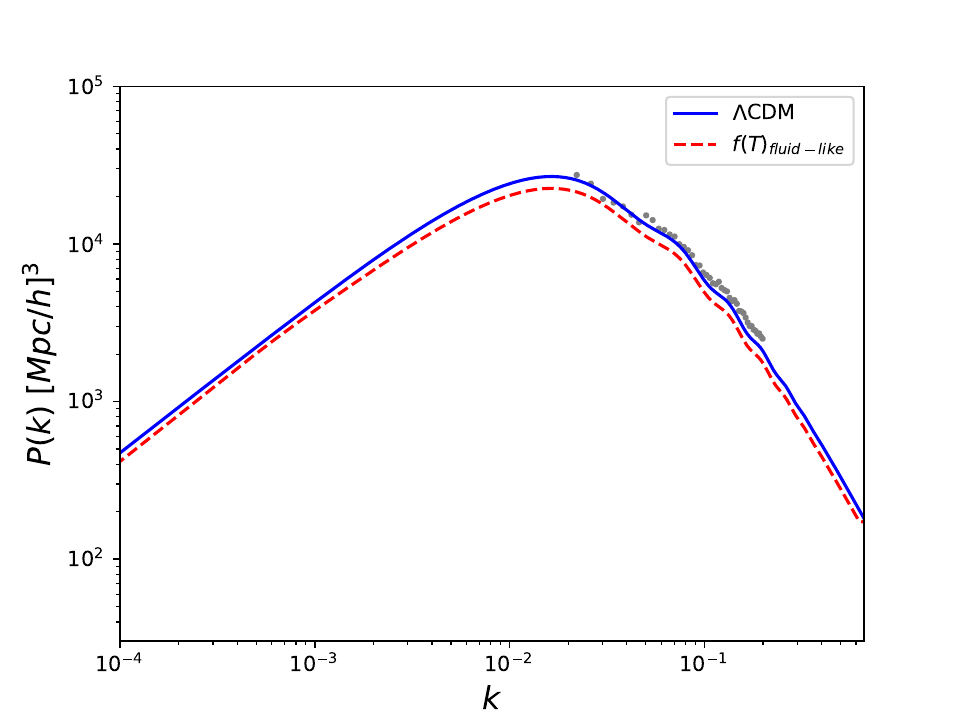}
\includegraphics[width=0.32\textwidth]{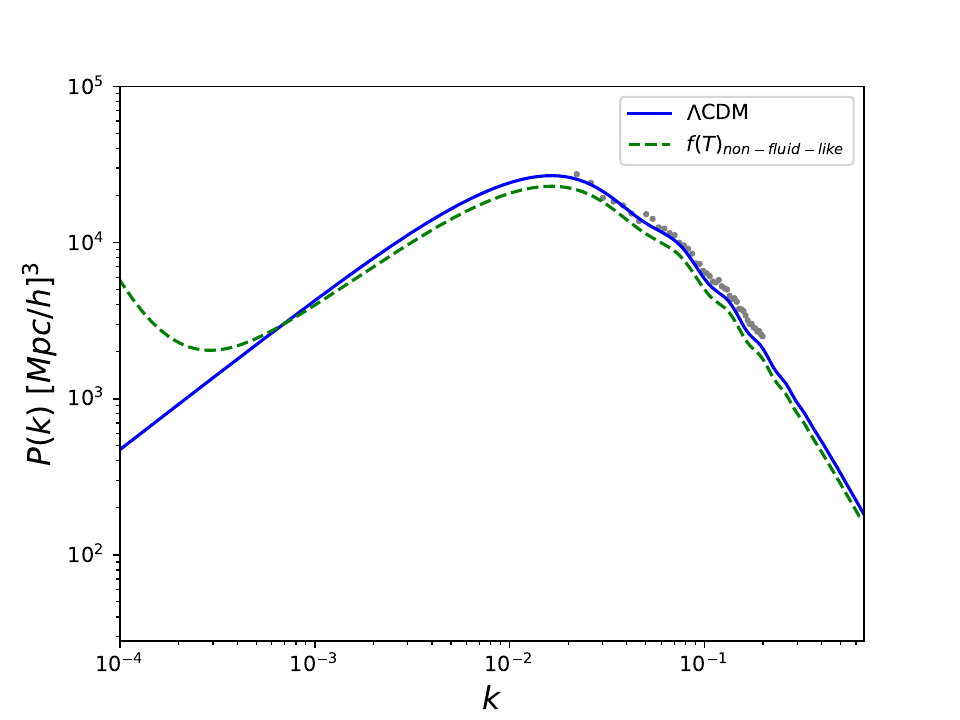}
\includegraphics[width=0.32\textwidth]{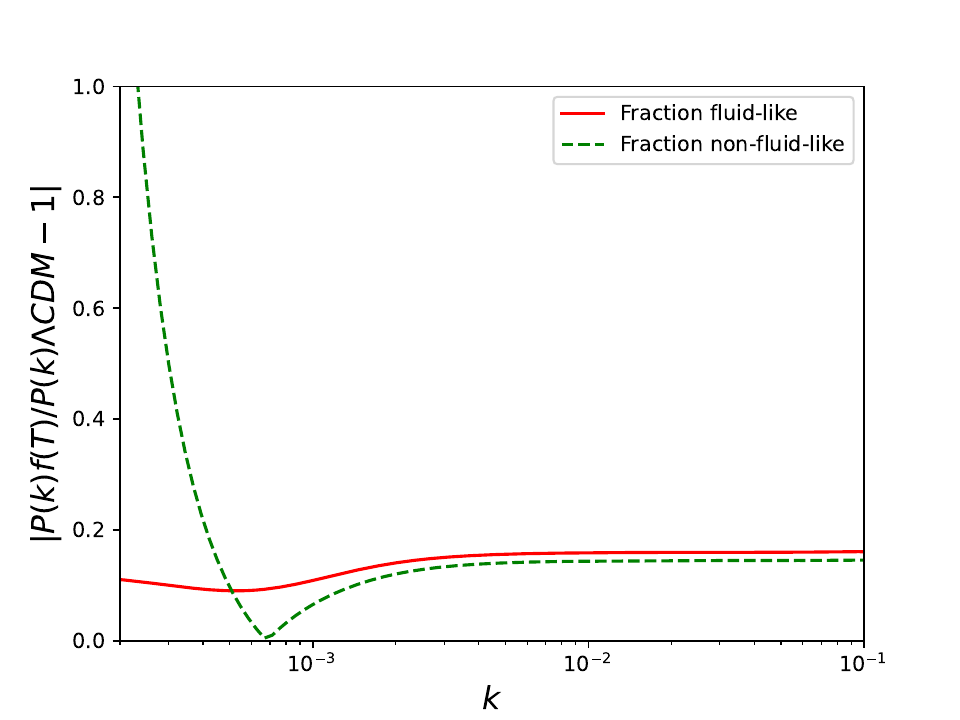}
\caption{Matter power spectrum for $f(T)$ power law model with $b=0.37$ (see Eq.~\eqref{PLM}) using the fluid-like approach (left) and non-fluid-like approach (middle). The $\Lambda$CDM model is also given (blue curve). In both of these plots, the grey dots correspond to SDSS DR7 LRG data. Furthermore, we show the percentage difference in the matter power spectrum between approaches (right). The green curve corresponds to the ratio of the non-fluid-like approach over $\Lambda$CDM, while the red curve is the ratio of the fluid-like approach over $\Lambda$CDM.}
\label{fig:MS}
\end{figure}

\textit{On the matter power spectrum.-} In Figure \ref{fig:MS} we show the matter power spectrum. In the left figure, we compare the fluid-like approach (red and dashed curve) with $\Lambda$CDM (blue and solid curve). The SDSS DR7 LRG likelihood was also included (grey points). In the middle figure, we present the non-fluid-like approach (green and dashed curve). Also, in the right figure we computed the difference between approaches using the expression $|\frac{{P(k)f(T)}}{{P(k)\Lambda}\text{CDM}}-1|$. From these three analyses, we can see that the non-fluid-like approach adopts higher values in the long wave region of the spatial frequency, in comparison to $\Lambda$CDM and the fluid-like approach; in fact, we can notice more than 90\% of difference between approaches. After recombination, the standard model and the two approaches using the power law $f(T)$ model describe the same behaviour, but the main differences lie in the primordial epoch. The non-fluid-like describes a younger universe. 

\begin{figure}[H]
 \centering
\includegraphics[width=0.32\textwidth]{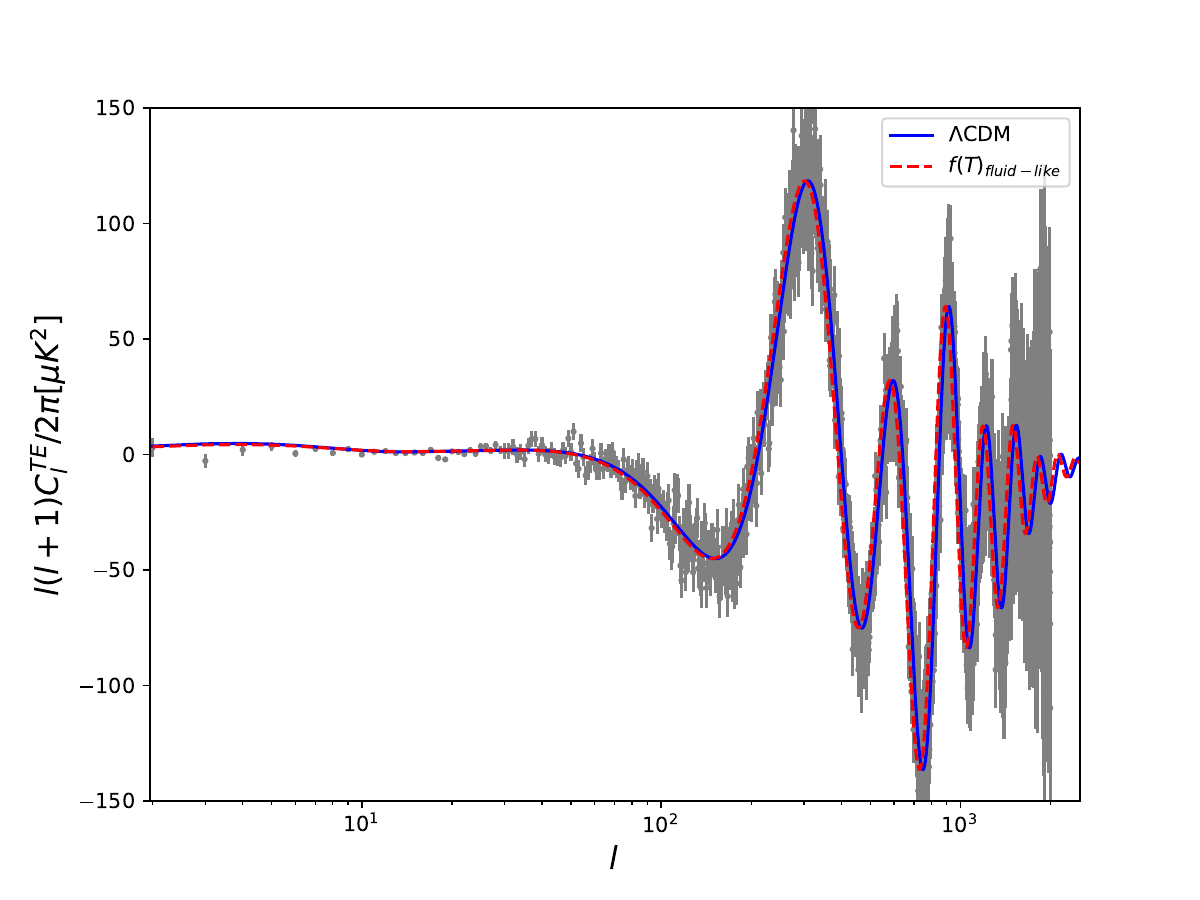}
\includegraphics[width=0.32\textwidth]{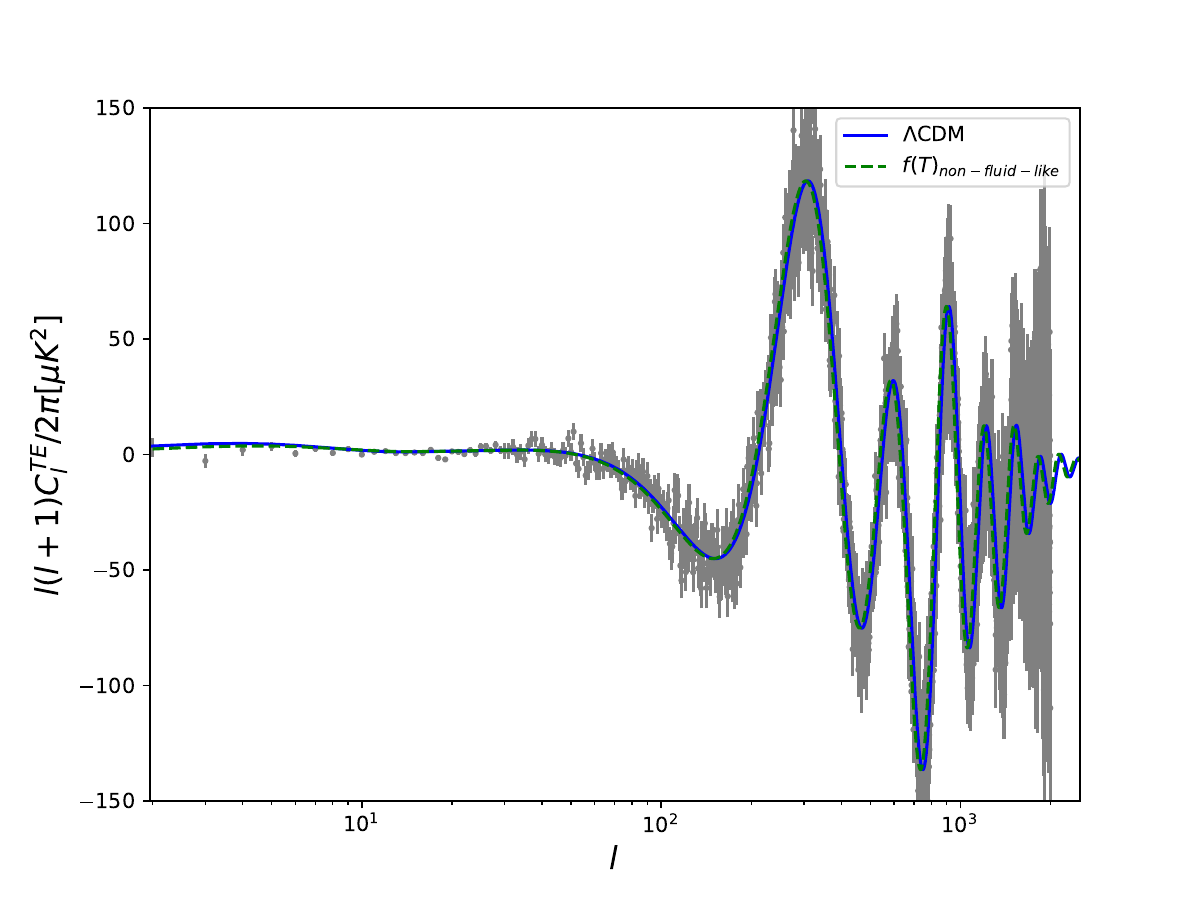}
\includegraphics[width=0.32\textwidth]{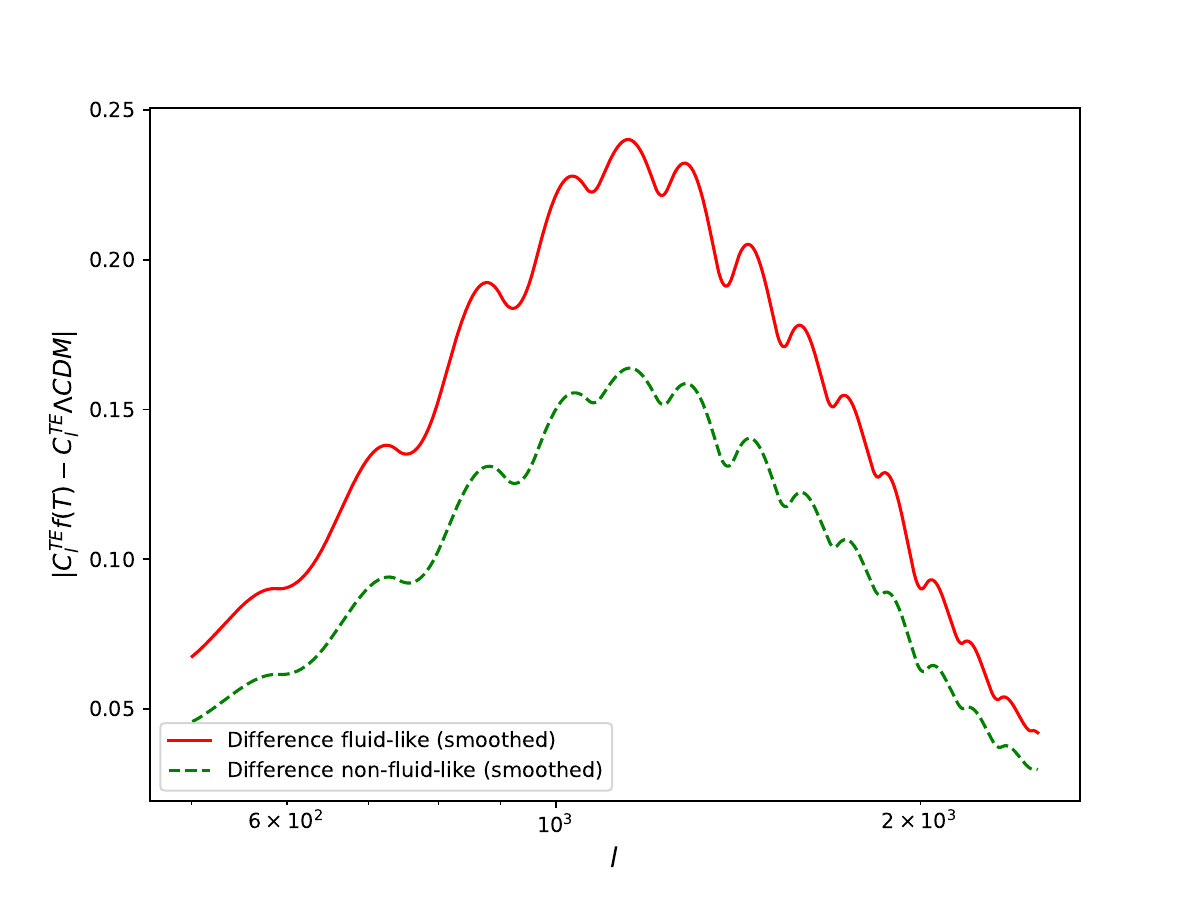}
 \caption{Fluid-like (left) and non-fluid-like (right) approaches power spectra.
 \textit{Top:} Temperature-Electric (TE) power spectrum for $\Lambda$CDM (blue curve) and for a $f(T)$ power law model with $b=0.37$ (see Eq.\eqref{PLM}). The grey dots and error bars correspond to Planck 2018 data \cite{plancklegacy}. 
The percentage difference (right) in the Temperature-Electric (TE) power spectrum is denoted by the green curve computing the subtraction of the non-fluid-like approach from the $\Lambda$CDM base model, while the red curve is the subtraction of the fluid-like approach from the $\Lambda$CDM base model. }
 \label{fig:TE}
\end{figure}

\textit{On the TE power spectrum.-} In Figure \ref{fig:TE}  we show the TE power spectrum. In the left figure, we compare the fluid-like approach (red and dashed curve) with $\Lambda$CDM (blue and solid curve). The TE CMB angular power spectrum dataset was also included in the plot (grey points and error bars). In the middle figure, we performed the non-fluid-like approach (green and dashed curve). We also computed the difference between approaches, as shown in the right figure, using the expression $|{C^{TE}_l}_{f(T)\text{approach}}-{C^{TE}_l}_{[\Lambda\text{CDM}]}|$. The evolution indicates that, even though the fluid-like approach deviates from $\Lambda$CDM from the non-fluid-like approach, the differences between both approaches are quite low. Therefore, the correlation between the temperature and electric fields for both approaches is not that different from the standard model.

\begin{figure}[H]
 \centering
\includegraphics[width=0.32\textwidth]{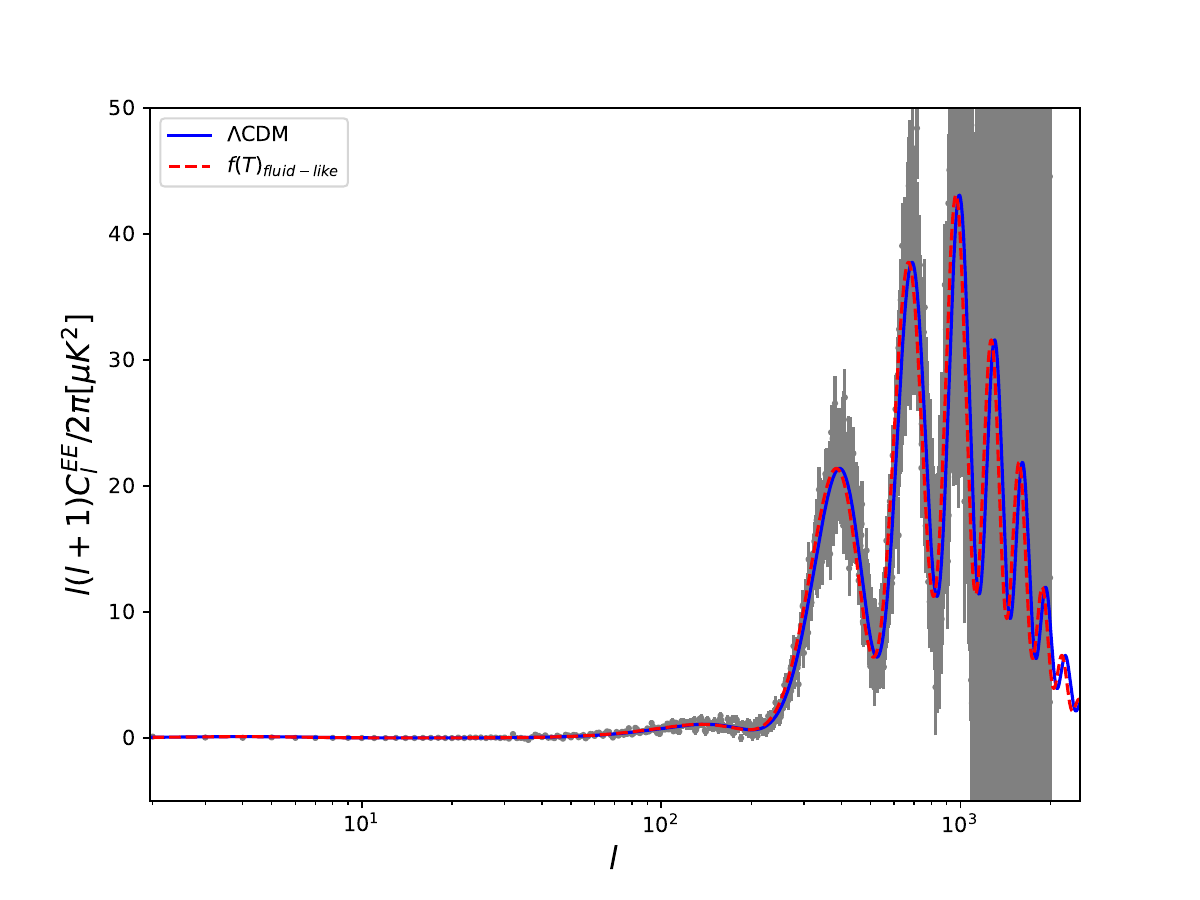}
 \includegraphics[width=0.32\textwidth]{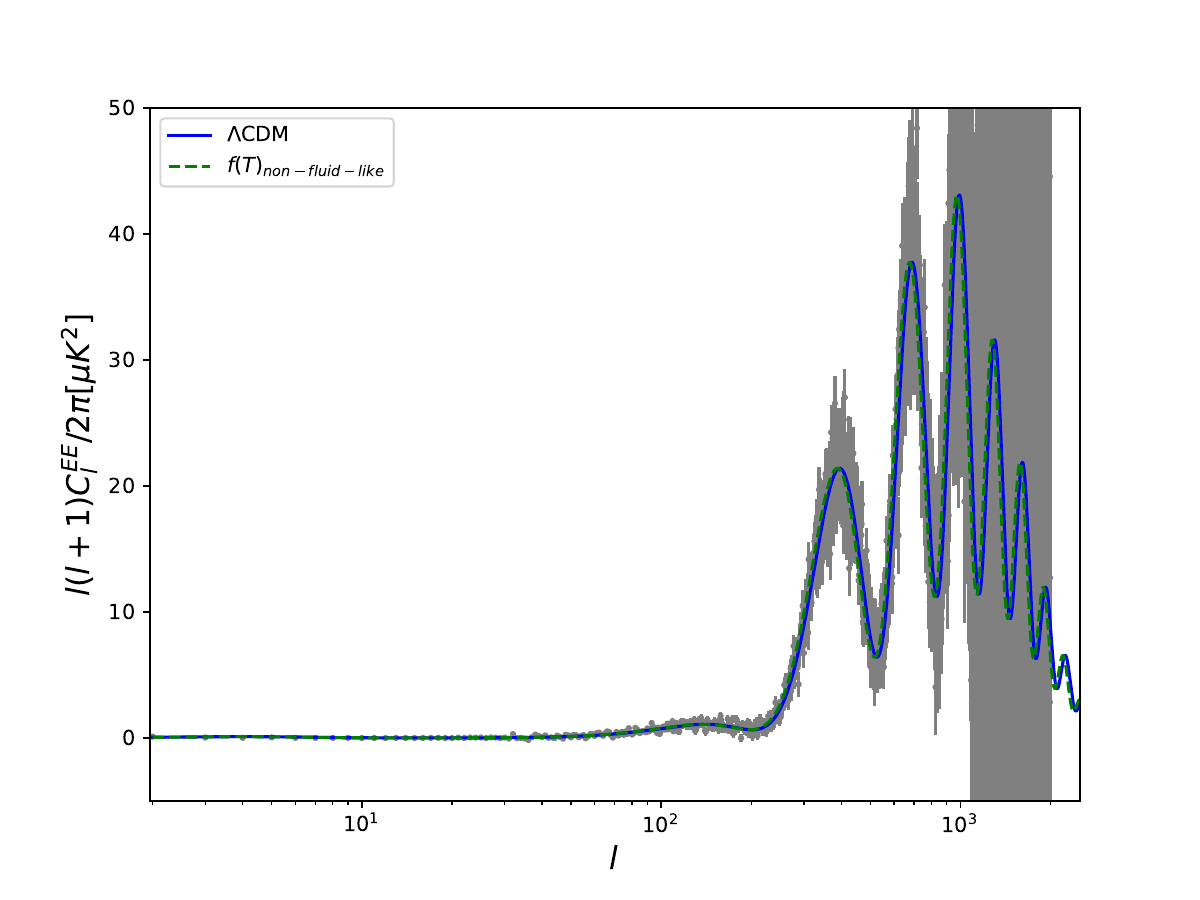}
 \includegraphics[width=0.32\textwidth]{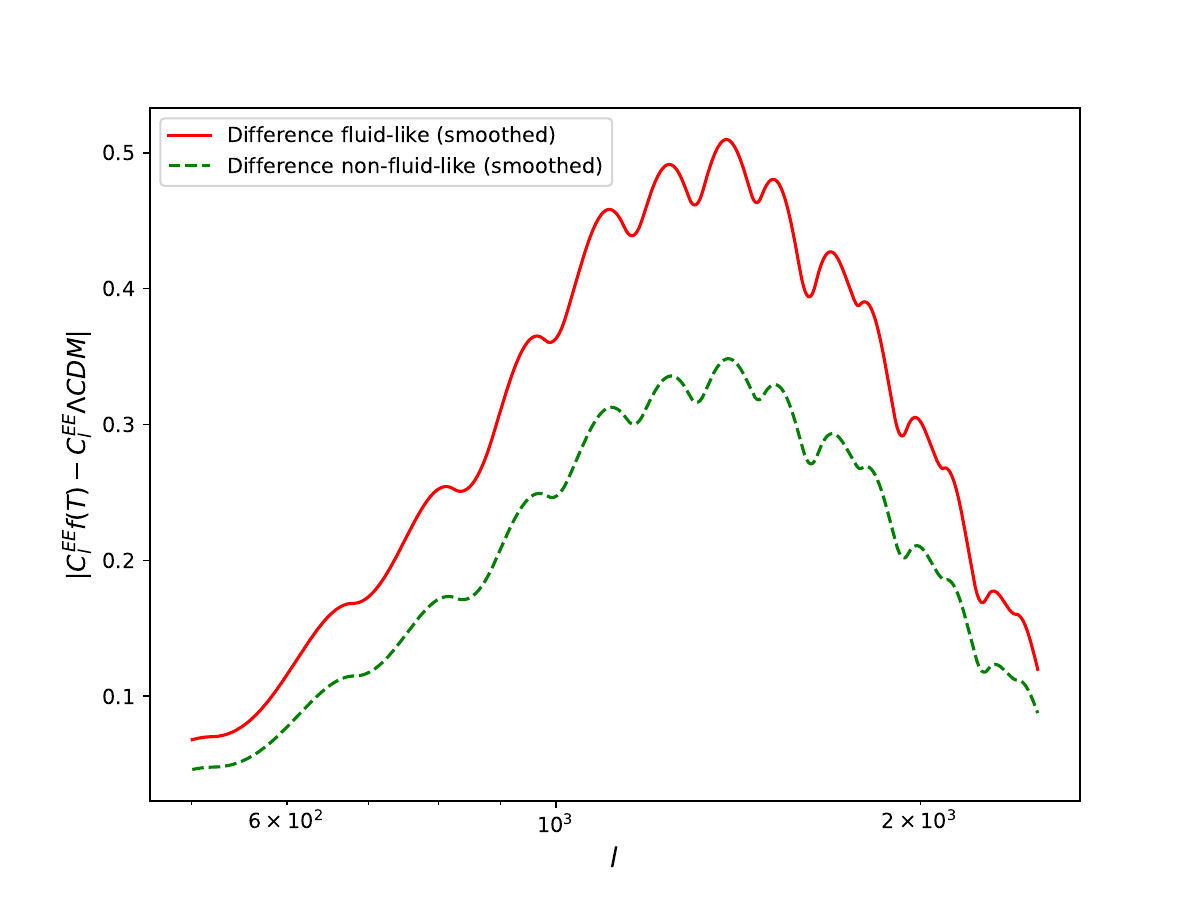}
 \caption{Electric (EE) power spectrum for $\Lambda$CDM (blue curve) and for a $f(T)$ power law model with $b=0.37$ (see Eq.\eqref{PLM}). The grey dots and error bars correspond to Planck 2018 data \cite{plancklegacy}. 
The percentage difference (right) in the Electric (EE) power spectrum is denoted by the green curve computing the subtraction of the non-fluid-like approach from the $\Lambda$CDM base model, while the red curve is the subtraction of the fluid-like approach from the $\Lambda$CDM base model. 
 }
 \label{fig:EE}
\end{figure}

\textit{On the EE power spectrum.-} In Figure \ref{fig:EE}  we show the EE power spectrum. In the left figure, we compare the fluid-like approach (red and dashed curve) with $\Lambda$CDM (blue and solid curve). The EE CMB angular power spectrum dataset was also included in the plot (grey points and error bars). In the middle figure, we did it for the non-fluid-like approach (green and dashed curve). We also computed the difference between approaches, as shown in the right figure, using the expression $|{C^{EE}_l}_{f(T)\text{approach}}-{C^{EE}_l}_{[\Lambda\text{CDM}]}|$. Similarly to the TE case, the fluid-like approach deviates from $\Lambda$CDM larger than the non-fluid-like approach, but the differences between both approaches are again low. The auto-correlation of the electric field for both approaches does not change from the standard model. 

\begin{figure}[H]
 \centering
\includegraphics[width=0.32\textwidth]{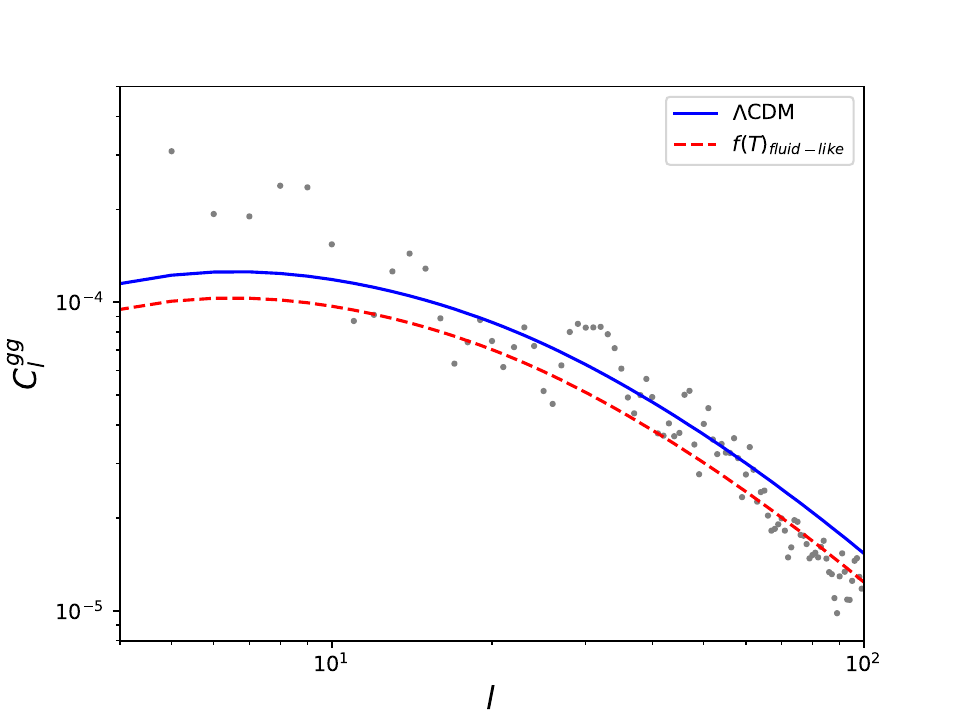}
\includegraphics[width=0.32\textwidth]{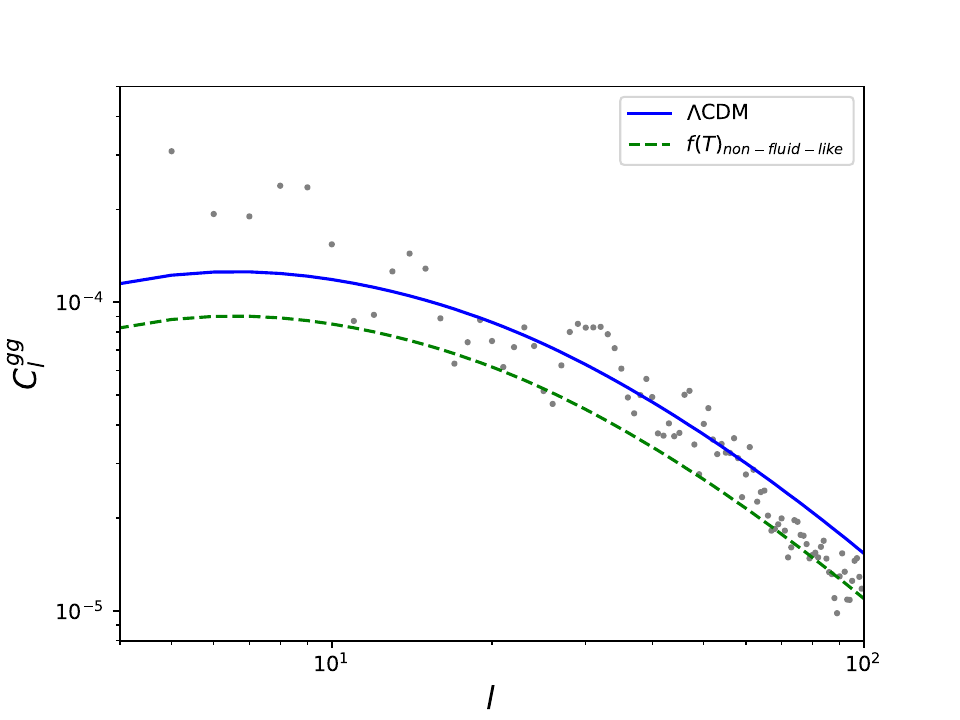}
 \includegraphics[width=0.32\textwidth]{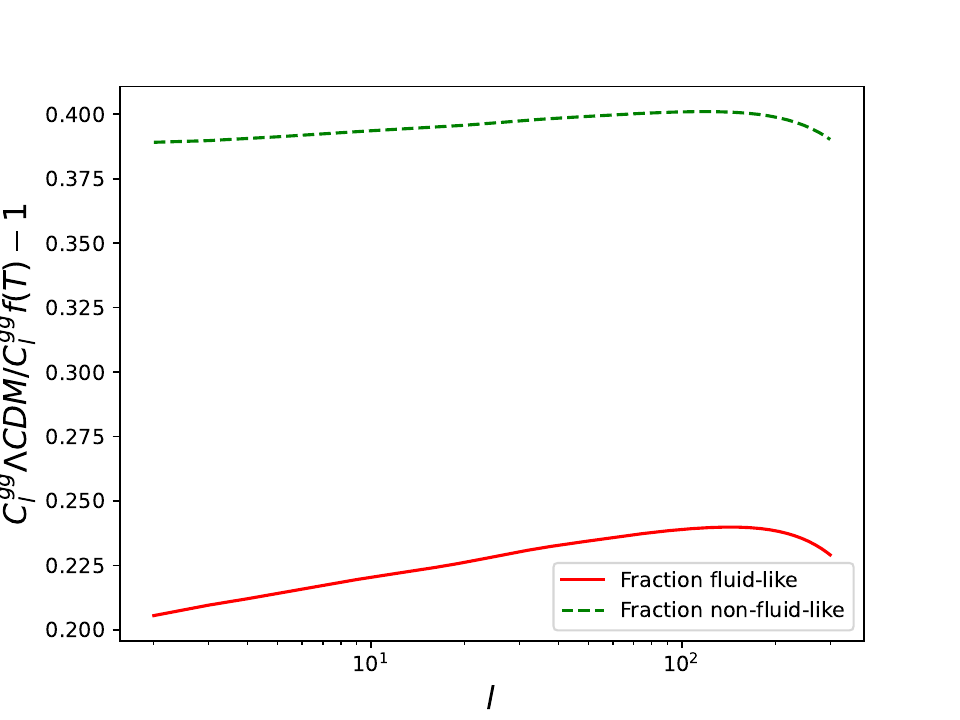}
 \caption{
Best fit shear-shear auto-correlation angular power spectrum $
 C^{gg}_l$ model using the fluid-like approach (left) and the non-fluid-like approach (middle), both with $b=0.37$,  in the bin $z=0.1-0.3$. The two analyses are compared against the $\Lambda$CDM model (blue line). The observational data points correspond to the ISW data from the SDSS-DR12 catalog. 
Also, we show the percentage difference (right) between the $\Lambda$CDM model and the fluid-like (red) and non-fluid-like (green) approaches. 
 }
\label{fig:Cgg-and-Differences}
\end{figure}

\textit{On the angular power spectrum of shear-shear auto-correlation.- } In Figure \ref{fig:Cgg-and-Differences} we obtain the angular power spectrum of shear-shear auto-correlation $C^{gg}_l$; the proportionality constant in the predicted $C_{l}^{gg}$ was found by searching the best fit with the corresponding data, following \cite{Stolzner:2017ged}. Under this idea, we can use the full equation given by $C_{l}^{gg} = ml^{n}C_{l}^{\phi\phi}$, where $C_{l}^{\phi\phi}$ is the lensing angular power spectrum, which after the fit we can compute the proportionality factor with $ml^{n}$, where $m=17,300$ and $n=2.6$.
We computed the differences between both approaches against the $\Lambda$CDM base model \cite{Kou:2023gyc}. In order to compute this difference, we employed $\frac{{C^{gg}_l}_{[\Lambda\text{CDM}]}}{{C^{gg}_l}_{f(T)\text{approach}}}-1$.


\section{Conclusions}
\label{sec:conclusions}

In this paper, we discuss the implementation of $f(T)$ perturbed field equations in a Boltzmann code-named CLASS, which is widely used for the precision cosmology community. We started with the most general perturbed tetrad field and proceeded with the calculations of the Bardeen potentials through the field equations. We considered the particular case when $f(T)$ is a power law model, where we only have one single free parameter. We noticed that the fluid-like approach, most commonly used in the literature, is not optimal to modify the \textit{background.c} module in the code, therefore we are required to implement the full description of the perturbed $f(T)$ field equations into the \textit{perturbations.c} module. When considering an effective dark energy fluid $f(T)$, the code modules require the correct Boltzmann perturbed equations by perturbing the tetrad field as in \eqref{Ptetrad}. Furthermore, we need to consider the perturbed components of the energy-momentum tensor given in \eqref{Ene-Mo00}-\eqref{Ene-Moij}. This approach will bring field equations which are different from the first approach. Following our proposal for a non-fluid like approach, we found that there is a significant difference in the spectra which have a physical effect on cosmic behaviour at early times when a power law $f(T)$ model is considered. 

Notice that for the TT power spectrum, both, the fluid-like (FL) and non-fluid-like (NFL) approaches move slightly to the left (low multipoles) with respect to (\textit{w.r.t}) $\Lambda$CDM. This means principally that the dark energy density increases for both models \textit{w.r.t} $\Lambda$CDM, however, this increase is higher in the case of the NFL approach. Consequently, the NFL predicts a higher amount of dark energy and, therefore, higher cosmic acceleration. Hence, the value of $H_{0}$ should be necessarily higher, which will relax the respective tension in comparison to early time values. Regarding the right plot of Figure \ref{fig:TT}, we can notice an approximately 20\% difference between approaches at low multipoles, being the NFL the one which shows more discrepancy \textit{w.r.t} $\Lambda$CDM.

Regarding the TE and EE power spectra, in both components, NFL and FL approaches move to the left slightly \textit{w.r.t} $\Lambda$CDM. The displacement of the evolution in TE and EE has to do with the effect of dark matter and photon interaction. Because NFL and FL approaches are shifted to the left it indicates that these models lead to lower growth of matter fluctuations than $\Lambda$CDM. However, due to this displacement, the FL deviates more from $\Lambda$CDM than the NFL, which means that the growth of matter fluctuations is less in the FL approach than in the NFL approach. In the case of the right plots of Figures \ref{fig:TE} and \ref{fig:EE}, we notice that, as it was mentioned before, the FL approach deviates more from $\Lambda$CDM rather than the NFL, that is the reason why the red curves are higher the green ones. In the case of the TE component, there is an approximate difference of 10\% between approaches, while in the EE component, a 20\% difference can be found.

In the case of the $P(k)$ matter spectrum, for greater $k's$, both approaches behave almost the same as the $\Lambda$CDM model. For lower $k's$, NFL grows highly \textit{w.r.t} $\Lambda$CDM and the FL approach. So, this means that in the primordial epoch, the NFL approach predicts lower matter density $\Omega_{m}$, while the FL predicts higher values. Moreover, NFL suggests a higher value of $\sigma_{8}$, while FL predicts a lower value. After recombination, the differences become smaller. NFL and FL are practically the same as $\Lambda$CDM. Hence, $\Lambda$CDM suggests a little more of $\Omega_{m}$ nowadays rather than NFL and FL. In the right plot of Figure \ref{fig:MS}, the percentage difference between approaches is higher under $k \sim 10^{-3}$, this is, at large scales. However, we notice a difference higher than 90\%, this could be related to a numerical defect. Since in both approaches we are using the same perturbative gauge and parameters, this is an important indicator that the two approaches are inequivalent at the primordial epoch.

At last aspect is related to the shear-shear power spectrum. The behaviour of $\Lambda$CDM, FL and NFL is the same but just shifted; FL is above NFL and under $\Lambda$CDM, and NFL is the one which adopts lower values. Since $C_{l}^{gg}$ describes the light deformation which arrives to us from galaxies due to scalar perturbations (presence of matter), then it means that NFL predicts less matter in the universe, which is consistent with the $P(k)$ matter spectrum results. This means that the $\sigma_{8}$ grows considerably in NFL \textit{w.r.t} $\Lambda$CDM and, therefore, in comparison to local observations, the $\sigma_{8}$ tension gets worse. As it was mentioned in \cite{Benetti:2020hxp}, the correlation between $H_{0}$ and $\sigma_{8}$ is not removed from $f(T)$ models. From the right plot of Figure \ref{fig:Cgg-and-Differences}, we can notice a 17\% difference between approaches. Once again, the NFL is the one which shows more discrepancy \textit{w.r.t} $\Lambda$CDM.

Regarding such results, we notice that a fluid-like approach could be a good first outlook to explore a background cosmology. However, at perturbative level, we expect that the non-fluid like approach proposed in this work will differ from the first one in terms of lowering cosmological tensions due to the numerical implementations in the thermodynamical equations. Further constraint analysis at statistical level will be reported elsewhere.


\begin{acknowledgments}
This research has been carried out using computational facilities procured through the Cosmostatistics National Group ICN UNAM project.
CE-R acknowledges the Royal Astronomical Society as FRAS 10147. 
A.A acknowledges financial support from SEP–CONACYT postgraduate grants program.
This article is based upon work from COST Action CA21136 Addressing observational tensions in cosmology with systematics and fundamental physics (CosmoVerse) supported by COST (European Cooperation in Science and Technology).
\end{acknowledgments}


\bibliographystyle{unsrt}
\bibliography{references}

\end{document}